\documentclass[review]{elsarticle}

\usepackage{lineno,hyperref}
\usepackage{graphicx}
\usepackage{subcaption}
\usepackage{epstopdf}
\usepackage{multirow}
\usepackage{array}
\newcolumntype{C}[1]{>{\centering\let\newline\\\arraybackslash\hspace{0pt}}m{#1}}
\usepackage{xcolor}
\journal{Journal of \LaTeX\ Templates}

\biboptions{authoryear}

\begin{document}

\begin{frontmatter}

\title{Extensive study of radiation dose on human body at aviation altitude
through Monte Carlo simulation}

\author[csp-address]{Abhijit Roy}

\author[csp-address]{Ritabrata Sarkar\corref{cor1}}
\ead{ritabrata.s@gmail.com}
\ead{aviatphysics@gmail.com}

\author[nci-address]{Choonsik Lee}

\cortext[cor1]{Corresponding author}

\address[csp-address]{Indian Centre for Space Physics, 43 Chalantika, Garia
Station Rd.,  Kolkata 700084, W.B., India}

\address[nci-address]{Division of Cancer Epidemiology and Genetics, National
Cancer Institute, National Institutes of Health, Bethesda, MD 20852, USA}

\begin{abstract}
The diverse near-Earth radiation environment due to cosmic rays and solar
radiation has direct impact on human civilization. In the present and
upcoming era of increasing air transfer, it is important to have precise idea of
radiation dose effects on human body during air travel. Here, we calculate the
radiation dose on the human body at the aviation altitude, also considering the
shielding effect of the aircraft structure, using Monte Carlo simulation
technique based on Geant4 toolkit. We consider proper 3D mathematical model of
the atmosphere and geomagnetic field, updated profile of the incoming particle
flux due to cosmic rays and appropriate physics processes. We use
quasi-realistic computational phantoms to replicate the human body (male/female)
for the effective dose calculation and develop a simplified mathematical model
of the aircraft (taking Boeing 777--200LR as reference) for the shielding study.
We simulate the radiation environment at the flying altitude (considering
geomagnetic latitude in the range of 45-50$^{\circ}$), as well as at various
locations inside the fuselage of the aircraft. Then, we calculate the dose rates
in the different organs for both male and female phantoms, based on latest
recommendations of International Commission on Radio logical Protection.
This calculation shows that the sex-averaged effective dose rate in human 
phantom is 5.46 $\mu$Sv/h, whereas, if we calculate weighted sum 
of equivalent dose contributions separately in female and male body: total 
weighted sum of equivalent dose rate received by the 
female phantom is 5.72 $\mu$Sv/h and that by the male phantom is 5.20 $\mu$Sv/h. 
From the simulation, we also calculate the numerous cosmogenic radionuclides 
produced inside the phantoms through activation or spallation processes which 
may induce long-term biological effects.
\end{abstract}


\end{frontmatter}

%\linenumbers

\section{Introduction}
\label{sec:intro}
Two major inputs controlling the near Earth space radiation environment are
Galactic Cosmic Radiation (GCR) and Solar Particle Events (SPEs). The particles
and radiation have a wide energy range spanning from few MeV to above $10^{12}$
GeV with decreasing flux as the energy increases. The nature and origin of these
particles depend on their energy. The low energy radiation observed below about
100 MeV are mainly from solar origin. However, during the SPEs \citep{SEP}
the energy of the solar particles or radiation due to coronal mass ejection or 
solar flares, can extend up to several GeVs and the flux can last from an hour 
to days \citep{gopa06}. On the other hand, GCRs of Galactic or extra-galactic 
origin, while propagating through the heliosphere, interact with the 
electromagnetic field carried by the solar energetic particles, which modifies 
the intensity of low energy GCR particles up to the energy of several GeVs \citep{gais90}. The 
heliospheric space-radiation environment is highly dynamic and strongly 
correlated with solar activities.

The intensity of space radiation near Earth depends on the strength and spatial 
distribution of Earth's magnetic field \citep{stor55}. The charged particle
(spectral and spatial) distribution is modified by the geomagnetic field
surrounding the Earth which also depends on the solar activity. There is a
strong dependence of the rigidity cut-off of the primary GCR and SPE particles 
on the geomagnetic latitude and solar activity \citep{bazi98, sark17}.
After entering the Earth's atmosphere, the radiation and particles interact
with the atmospheric molecules, producing a large number of secondary particles
in cascade, which is known as extensive air shower. Due to the balance between
the production rate of the secondary particles and their absorption in the
atmosphere, the particle flux has a maximum at around 15-20 km altitude (also
depends on the geomagnetic latitude), which is called the Regener-Pfotzer 
maximum \citep{rege35}. The particle flux starts to decrease below this region.

The radiation environment in the atmosphere due to the primary Cosmic Ray (CR)
particles can produce serious biological hazard to high altitude travelers and
also can disrupt the microelectronics components in the aircraft. The radiation
intensity and hence the induced potential hazard generally increases with 
altitude and geomagnetic latitude. The CR particles and radiation have enough
energy to give a serious amount of radiation dose on human body which may lead
to fatal cancer, eye cataracts, cardiovascular diseases, malfunctions of the
central nervous system etc. From the study of \citet{benn13} on Air Canada
pilots shows that, most of them received annual dose about 3 mSv even a few of
them received approx 5 mSv, which is higher than the International Commission 
on Radiological Protection (ICRP) recommended 1 mSv per year limit for public 
radiation exposure. Due to this high exposure from CR on flight crews, disorder 
in the reproductive system and high risk of miscarriage among female flight 
attendants was also observed \citep{graj15, laur06, asph99}. For these reasons 
pilots and flight crews are considered as radiation workers by ICRP 
\citep{icrp07}. Continuous exposure to this kind of radiation can also  produce 
various radioactive isotopes in our body \citep{brod69}, where these could 
cumulatively give radiation dose depending upon their half lives.

There are different protection and operational quantities used in radiation
dosimetry like: \textit{equivalent dose} (H$_{T}$) (in different organ type T),
\textit{effective dose} (E) and \textit{ambient dose equivalent} (H$^{*}$(10)).
The definition of these quantities can be found in ICRP publication 103
\citep{icrp07}. As the protection quantities like E, H$_{T}$ cannot be measured
directly in the human body, so the operational quantity (H$^{*}$(10)) is used
for measurements, from which the protection quantities can be derived using
suitable conversion coefficient. However, in many cases ambient dose equivalent
cannot properly estimate the effective dose \citep{ferr97, sato99}.

There are different computer programs based on Monte Carlo calculation or
empirical models to calculate radiation dose at flight altitude \citep{bott12}.
But there are significant number of uncertainties associated with these kind
of calculations which is summarized in \cite{icrp10} and EURADOS report
\citep{euro04}. Also, there are a limited number of studies that consider the
effect of aircraft structure for dose calculation \citep{batt05, ferr04, dyer03,
dyer01}. These calculations also have several limitations associated with them 
like, considering the isotropic distribution for fluence to dose calculation 
and fixed vertical cut-off rigidity instead of direction dependence of cut-off 
rigidity.

There is no experimental or calculated radiation dose data at individual organs
of flight crews and a few study on the shielding effect of aircraft structure.
These studies can be important for planning the radiological protection. In this
concern, we develop a computer simulation program based on Geant4 \citep{GEANT4}
Monte Carlo simulation toolkit to calculate the absorbed dose rate, equivalent
dose rate in individual organs, organ weighted sum of 
equivalent dose rate for both male and female body as well as the sex-average
effective dose rate at aviation altitude radiation environment. The change in
particle fluxes inside the aircraft due to the shielding effects of the aircraft
structure also have been studied here considering full-scale aircraft model.

In the following Sec. \ref{sec:simu}, we describe the simulation method for the
calculation of radiation dose at the aviation altitude. We present and discuss
the simulation results in Sec. \ref{sec:result}. And finally conclude in Sec.
\ref{sec:conc}.

\section{Simulation Method for Dose Calculation}
\label{sec:simu}
The overall simulation of radiation dose at flight altitude comprises a series
of calculations. First, we calculated the radiation environment at the relevant
altitude for which we need to consider proper models for the Earth's
atmosphere, geomagnetic field and distribution of the primary GCR. To simulate 
the radiation environment inside the aircraft, we need to consider a proper 
structure of the aircraft. Similarly, to calculate the radiation dose in human 
body, proper computational description of the human phantom is required. We 
also need to consider the underlying physics processes properly to define the
radiation interactions with matter.

\subsection{Simulation of secondary radiation environment in atmosphere}
\label{ssec:srad}
The secondary CR radiation environment at the Earth's atmosphere is formed due 
to the interaction of primary GCR with Earth's atmosphere. Here, in this work 
we did not consider the sporadic effects due to the high energy particles from 
SPE. A detailed description of the simulation procedure for primary GCR 
interaction with the Earth's atmosphere to produce secondary radiation and 
particle fluxes can be found in \citet{sark20}. For this simulation purpose, 
we considered full 3D atmospheric and magnetospheric models and with updated 
Local Interstellar Spectrum (LIS) \citep{herb17} for the primary GCR. We used NRLMSIS-00 
\citep{pico02} atmospheric model to define the Earth's atmosphere up to 100 km 
altitude from the surface subdivided into 100 concentric shells with thickness 
equal in logarithmic scale. The Earth's magnetosphere was defined using: 
(i) 12th generation IGRF model \citep{igrf15} for the inner magnetic field 
distribution and (ii) Tsyganenko Model \citep{tsyg16} for the external magnetic 
field distribution, with proper input parameters.

A proper description of primary GCR LIS is very important in order to simulate 
the shower of secondary particles in the Earth's atmosphere. At the end of 
2012, for the first time {\it in situ} measurement of unmodulated primary GCR 
was possible by \textit{Voyager1} \citep{ston13} after crossing the heliopause. 
Complimenting the PAMELA \citep{adri13} and AMS-02 \citep{agui15} measurements 
with the \textit{Voyager1} measurement, a new model of very LIS, before 
modulation by the heliospheric electromagnetic field, was proposed by
\citet{vose15}. This LIS model was used to derive the differential GCR flux at 
1 AU by \citet{herb17} based on forced-field approximation \citep{caba04, 
mora13}. We considered this model to describe the GCR flux at the top of the 
atmosphere with an arbitrary value for the solar modulation potential $\phi$ = 
524 MV, (which corresponds to a solar transitional phase from maximum to 
nearing a minimal activity). 

\begin{figure}
\centering
\includegraphics[width=1.0\textwidth]{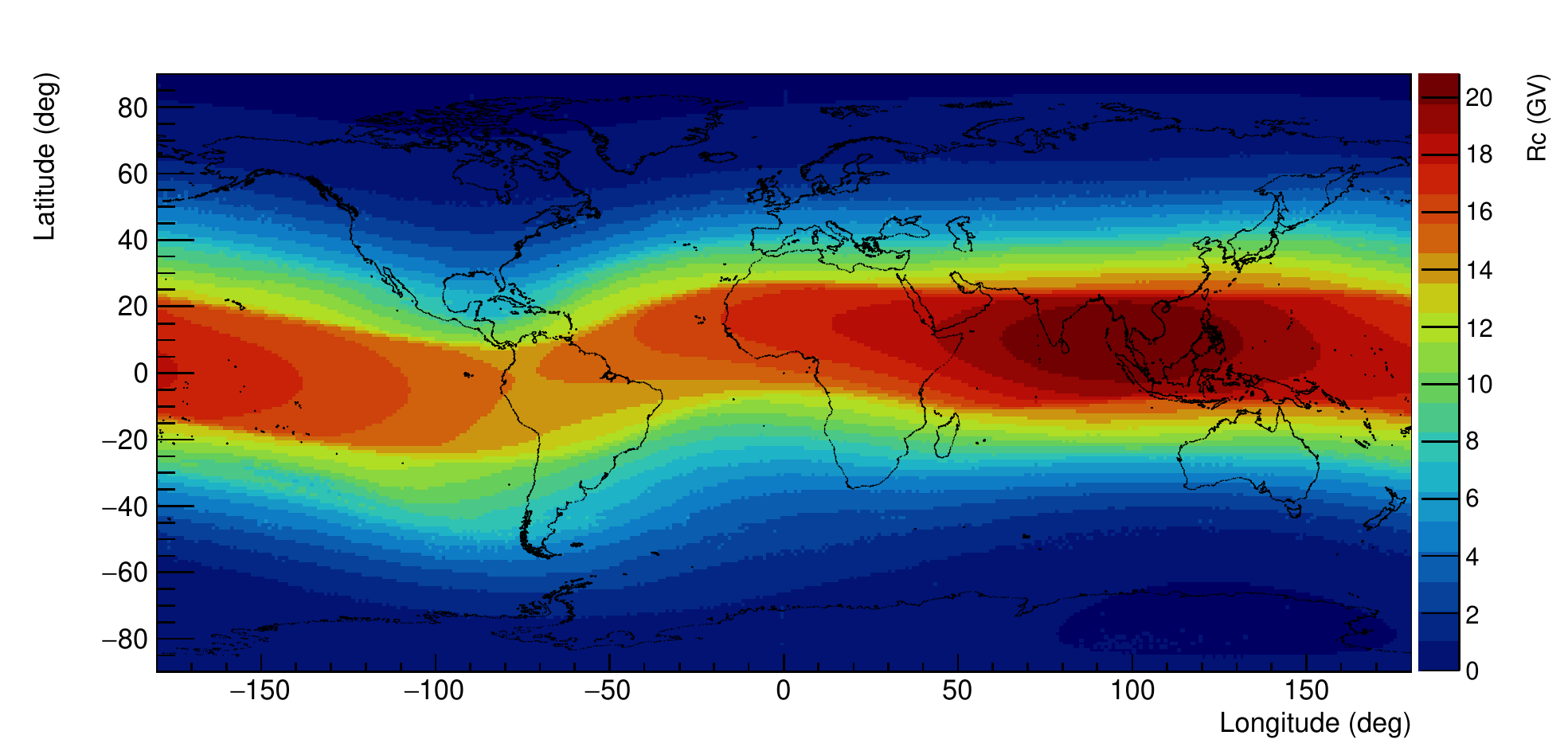}
\caption{Distribution of vertical rigidity cut-off at 10 km altitude over the total
geographic region.}
\label{fig:rcCut}
\end{figure}

The rigidity cut-off for the CR charged particles imposed by the geomagnetic
field distribution is an important factor for the calculation of secondary
radiation environment in the atmosphere. While, most of the other similar
program consider the vertical rigidity cut-off for the distribution of the
primary GCR at different locations, we considered the back-tracing method to
select the allowed tracks into the atmosphere from the isotropic distribution 
of particles at the top of the atmosphere \citep{sark20}. This method takes 
care of the directional dependence and penumbra region of the rigidity cut-off. 
However, for an example, Fig. \ref{fig:rcCut} exhibits the positional 
distribution of vertical rigidity cut-off at 10 km altitude. Due to high 
rigidity cut-off at the geomagnetic equator, only the charged particles with 
enough rigidity can penetrate the Earth's magnetic field and interact with the 
atmospheric nuclei, while the neutral particles are unaffected. 

\begin{figure}
\centering
\begin{subfigure} [b] {0.49\linewidth}
\includegraphics[scale=0.30]{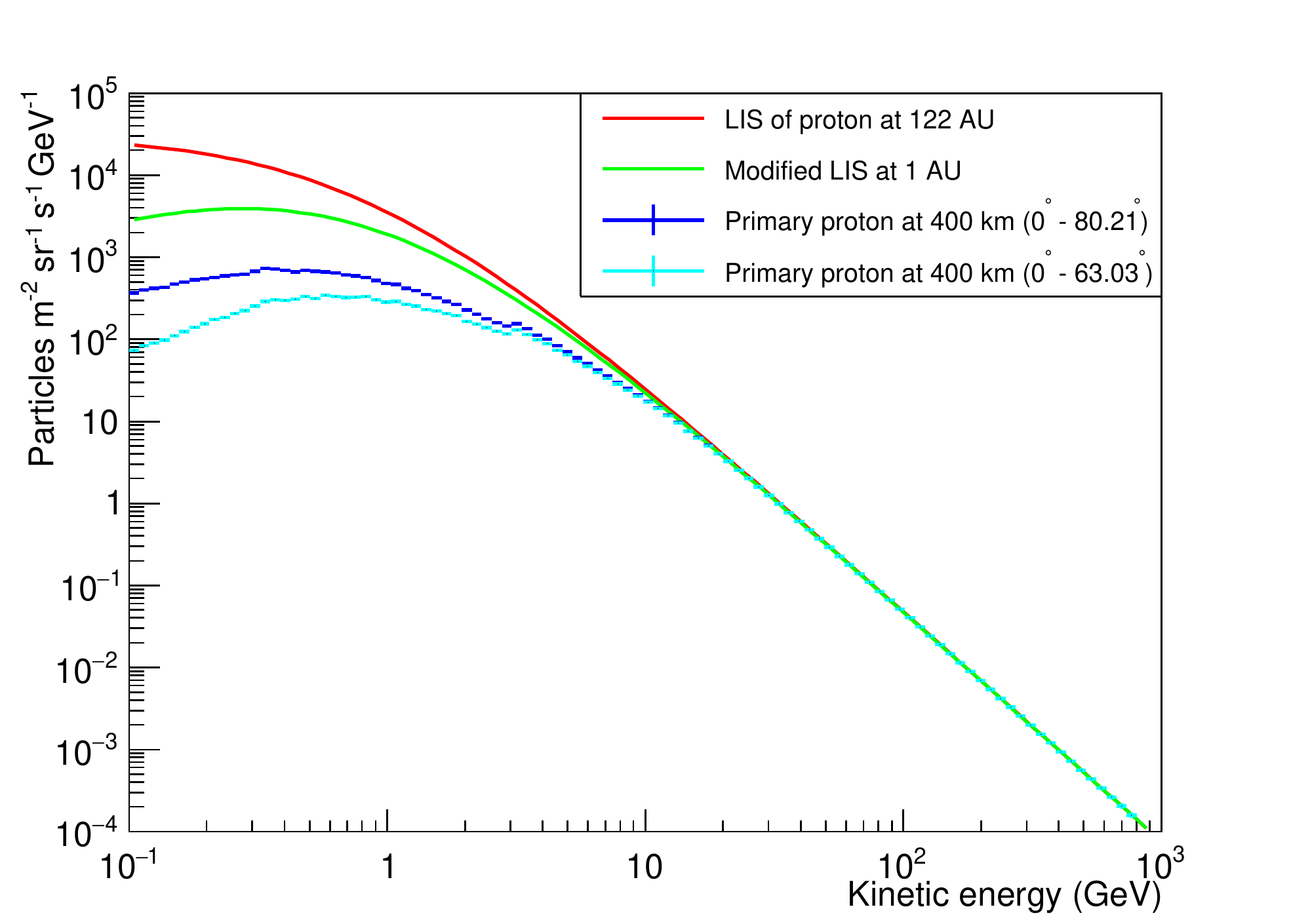}
\caption{Proton}
\end{subfigure}
\begin{subfigure} [b] {0.49\linewidth}
\includegraphics[scale=0.30]{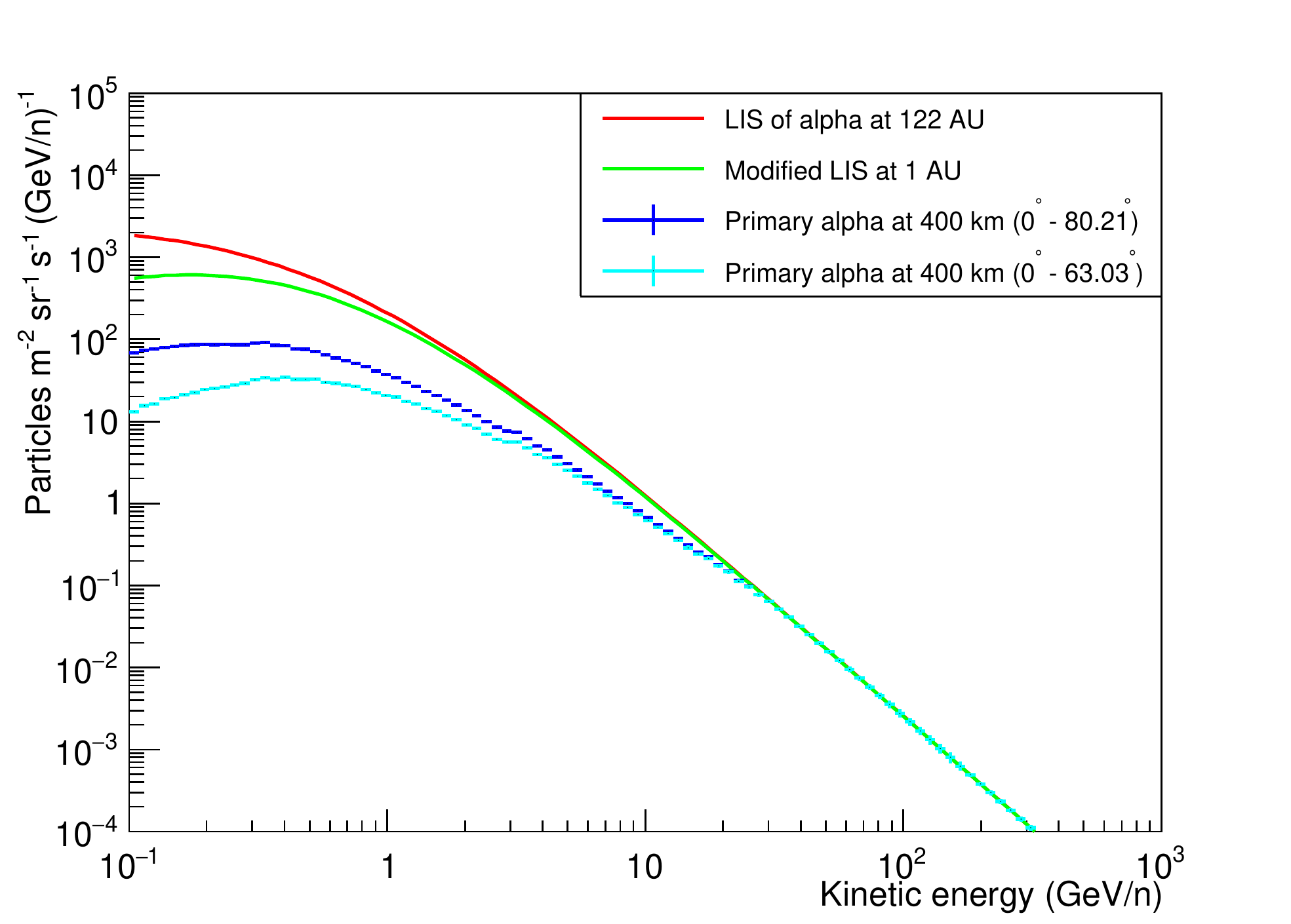}
\caption{Alpha}
\end{subfigure}
\caption{Very LIS of primary GCR in interstellar medium (red); modulated primary
GCR in heliosphere at 1 AU (green); modulated primary GCR by magnetosphere at
400 km in two different geomagnetic latitude range (blue, cyan).}
\label{fig:LIS}
\end{figure}

Using the geomagnetic field model and back-tracing method, the position
dependent geomagnetic modulation was inherently achieved in the simulation
as can be seen from Fig. \ref{fig:LIS}. This plot shows primary GCR spectra 
(for protons and alpha particles) at an altitude of 400 km, averaged over two
different geomagnetic latitude ranges 0-63.03$^{\circ}$ and 0-80.21$^{\circ}$,
along with the very LIS at 122 AU and modified LIS at 1 AU (modulated by the
electromagnetic field carried by the solar wind). We recorded all the relevant
information like: position, momentum, energy, type (pdg number) etc. of the 
generated secondary particles and radiation in this step of the simulation to 
be used afterward.

\subsection{Human phantom and aircraft model}
\label{ssec:massmod}
In the next stage of the calculation, we simulated the interaction of secondary 
radiations produced in the atmosphere (along with the primaries survived the
geomagnetic cut-off) in the human body and with the aircraft to see the
structural effects on radiation dose. 

The simulation of radiation dose received by human organs has been done
considering UF/NCI hybrid phantom models (PHANTOMS library), jointly developed
by the University of Florida and National Cancer Institute. These computational
models contain more than 100 organs or tissues. More details about PHANTOMS
and their compositions can be found in \citet{lee09,grif19} and references
therein. We can calculate the energy depositions separately in each of the
organs and subsequently get the overall effective dose in human body.

\begin{figure}
\centering
\includegraphics[width=1.1\textwidth]{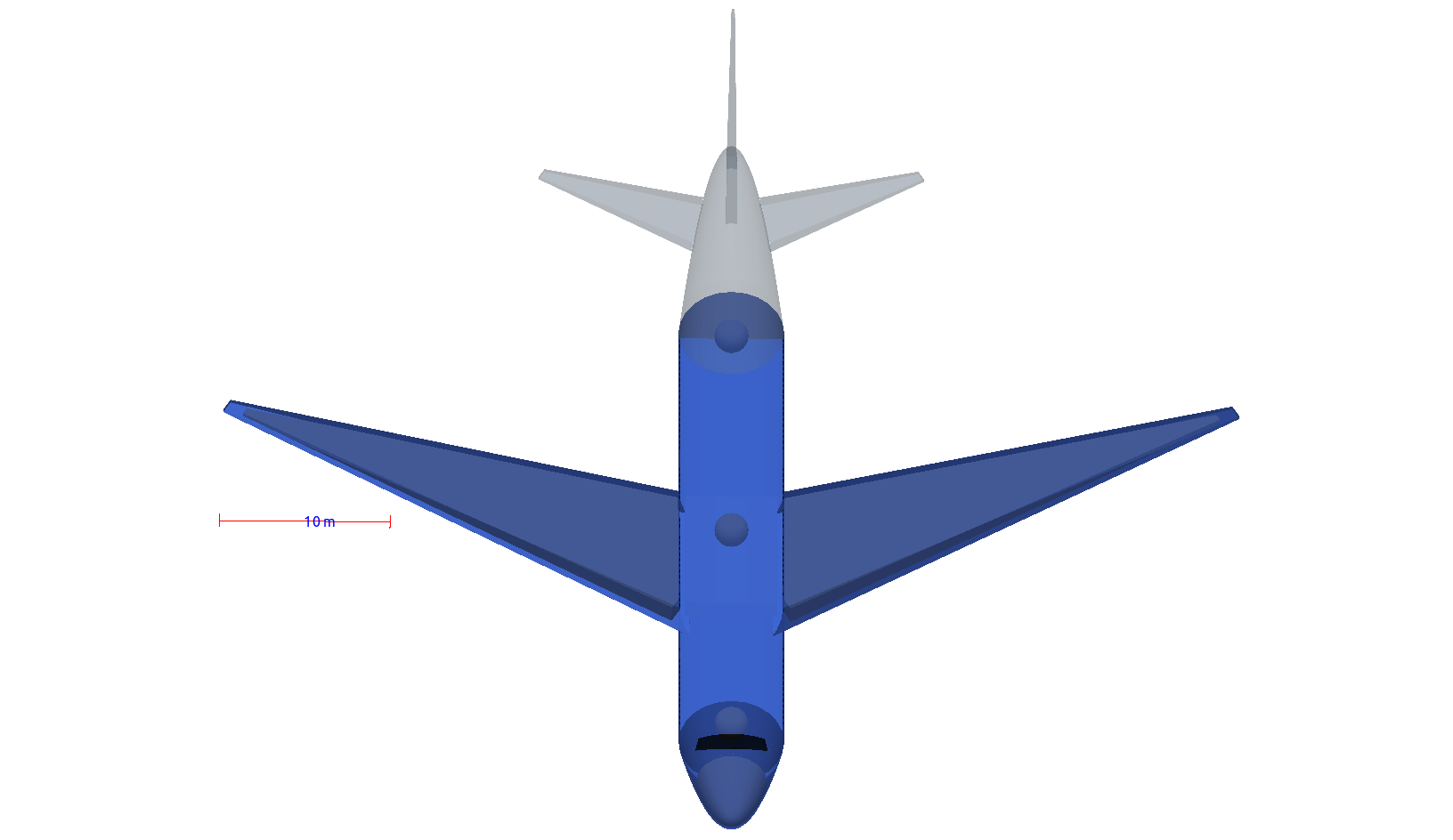}
\caption{Mass model structure of the aircraft used in the simulation with three
sampling dummy spheres placed at different sections inside the fuselage.}
\label{fig:geomAP}
\end{figure}

To find out the variation of radiation fluxes inside the aircraft structure, we
developed a simplified mass model of the aircraft structure in Geant4 
simulation framework, based on Boeing 777--200LR \citep{boei15} aircraft model. 
We sampled the radiation effects at different positions inside the aircraft by 
placing dummy volumes (three spheres of 2 m diameter at the front, middle and 
back sections of the fuselage). A representative picture of the aircraft model 
along with the dummy spheres can be seen in Fig. \ref{fig:geomAP}. A simplified
description of the jet fuel, window glass, cockpit instrumentation has been
considered. For the aircraft structure construction, we mainly used 5 mm thick
aluminum (1.35 g/cm$^2$) and the composition and density of other different 
materials (jet fuel, window glass, cockpit instrumentation) were taken from 
\cite{ferr04}. The full tanks of fuel (145,000 kg) were placed inside the wings 
and lower half of the fuselage confined along only in the region where the 
wings are connected to the fuselage. An additional aluminum plate of 5 mm 
thickness comprised the floor of the passenger cabin, thus rendering a total 
equivalent thickness of 2.7 g/cm$^2$ for the upward going particles.

\subsection{Physics list considered in the simulation}
\label{ssec:phylist}
It is important to consider the right set of physics processes to describe the 
radiation interactions with the atmosphere, aircraft structure and human 
phantom. For this simulation, we considered {\it QGSP\_BIC\_AllHP} physics list
\citep{geant4ph} with updated {\it TALYS-based Evaluated Nuclear Data Library
(TENDL)}, which is appropriate for hadronic interaction and isotope productions.
To handle electromagnetic interactions, {\it electromagnetic\_options3} physics
list was used, which is suitable for space and medical purposes.

\subsection{Simulation of radiation interaction in human phantom and aircraft
structure}
\label{ssec:phancraft}
From the knowledge of radiation environment at the flight altitude, we
simulated the radiation interactions in the human phantoms and the aircraft
structure. First, to calculate the radiation effect in human body we irradiated
the phantoms with different secondary GCR particles from upper and lower
hemispheres of 2 m radius (without considering the aircraft structure) for
both upward and downward going particles and calculated the corresponding
deposited doses. We used the standing posture of the phantoms for irradiation,
since there is no significant difference in radiation doses observed in the
sitting and standing postures \citep{alve16}.

To find out the variation of radiation inside the aircraft due to shielding
effects of aircraft structure, we irradiated the aircraft from two hemispheres
with 35 m radius along with three dummy spheres inside the fuselage at different
locations shown in Fig. \ref{fig:geomAP} and recorded the information of each
event inside those spheres.

For the purpose of radiation dose calculation in human body, we irradiated both
male and female phantoms by 10$^{6}$ number of each different types of
particles; while for the simulation of radiation interactions in the aircraft, 
we considered 10$^{7}$ particles from both upper and lower hemispheres. For
both simulations, we considered the particles with kinetic energy distribution
in the range of 10 keV to 800 GeV for downward going particles, while for upward
going particles the energy range is 10 keV to 10 GeV. This choice of energy
ranges can be justified by the simulated outcome of the secondary particle
fluxes in the atmosphere discussed later in Sec. \ref{ssec:fltalt}.

\section{Results and Discussions}
\label{sec:result}

\subsection{Radiation at flight altitude}
\label{ssec:fltalt}
\begin{figure}
\centering
\begin{subfigure} [b] {0.49\linewidth}
\includegraphics[scale=0.30]{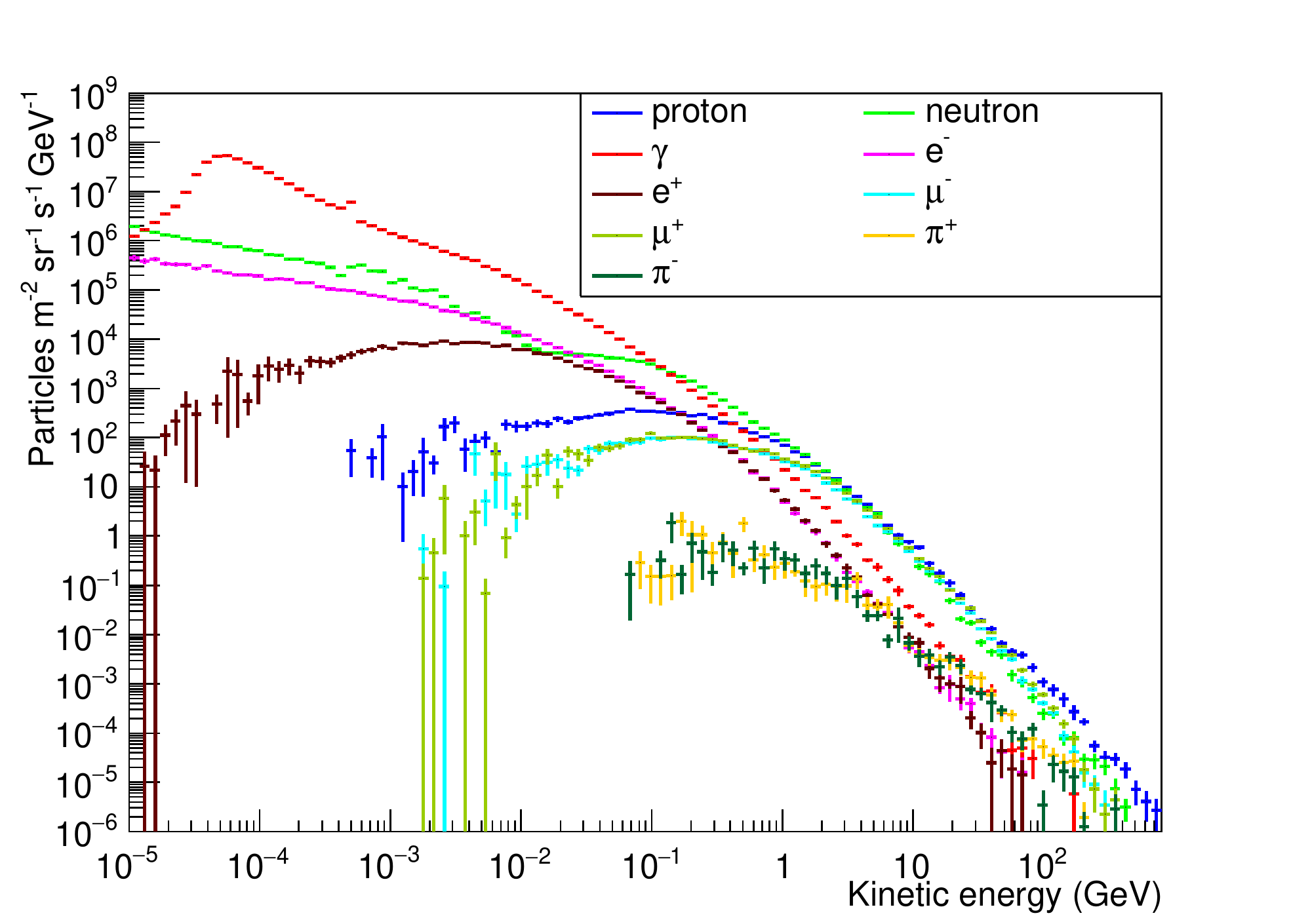}
\caption{Downward}
\end{subfigure}
\begin{subfigure} [b] {0.49\linewidth}
\includegraphics[scale=0.30]{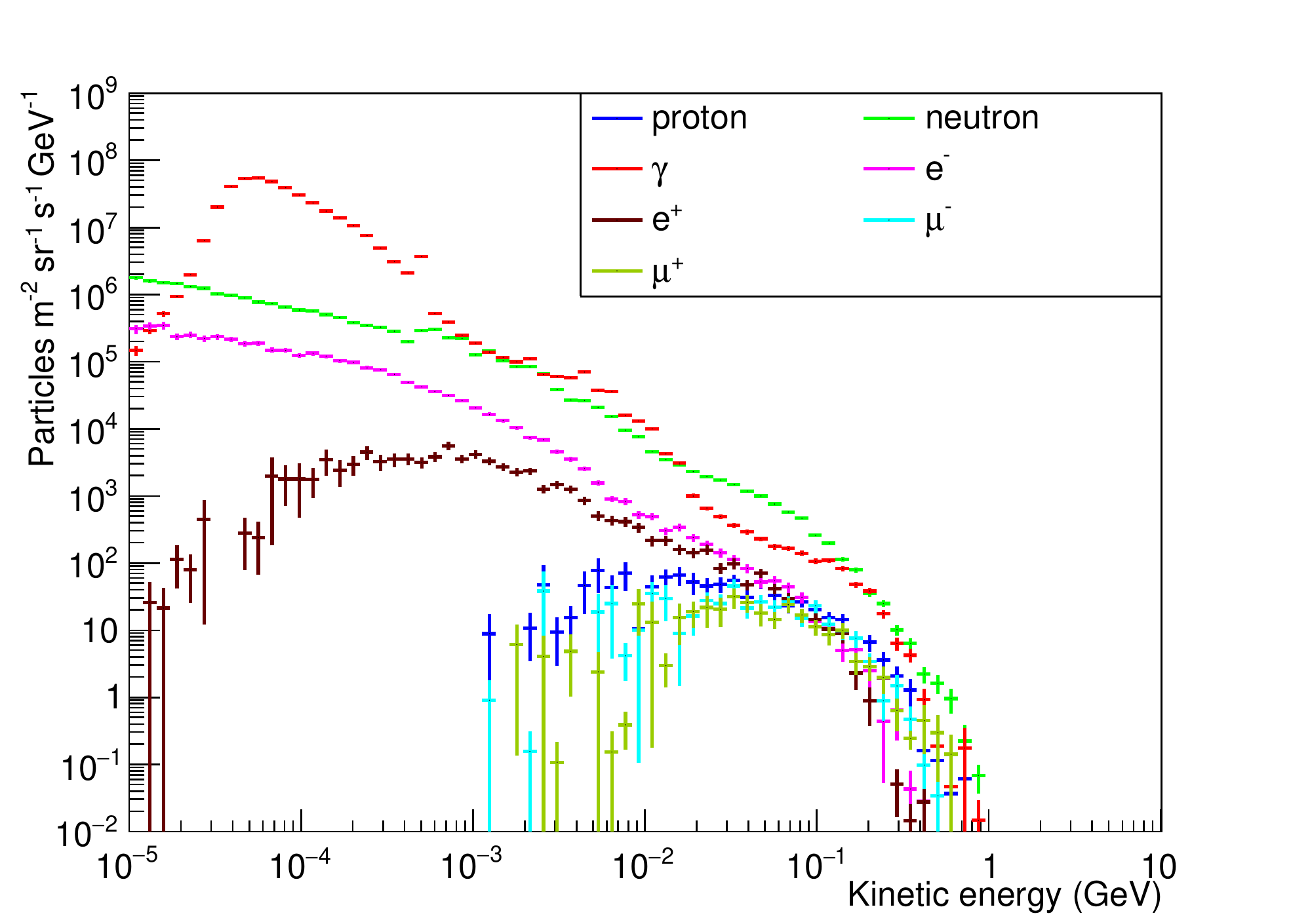}
\caption{Upward}
\end{subfigure}
\caption{Different secondary CR particle fluxes at 10 km altitude in
the geomagnetic latitude range of $\sim$45$^\circ$--50$^\circ$ for both
downward and upward going particles with respect to local zenith.}
\label{fig:seconEngFlux}
\end{figure}

The primary GCR LIS is modified in the heliosphere, by the electromagnetic 
field carried by the solar wind. At Earth's vicinity, this modified GCR is 
further modulated by the Earth's magnetic field, while propagating towards 
Earth through the magnetosphere. The primary particles with modified flux then
interact with the atmospheric nuclei to produce secondary particles and
radiation which in turn interact with the aircraft structure. The flux
distribution of all the different particles generated from the simulation at 
the flight altitude (which we considered here at 10 km from the Earth's 
surface) is shown in Fig. \ref{fig:seconEngFlux}. Though, due to the lack of 
direct experimental measurements of the distribution of radiation flux 
components at aviation altitude, it is not possible to directly verify the 
results shown in this figure. Nevertheless, we compared the simulation results 
at satellite altitude with AMS observation of proton flux and at balloon 
altitude with the atmospheric radiation measurement by ICSP balloon borne 
experiment \citep{sark20}. \cite{ferr05} mentioned the anisotropy in the 
distribution of high energy radiation flux components for downward and upward 
particles at aviation altitude, while low energy neutrons and photons $<$ 10 
MeV are more isotropically distributed. This fact is also evident from Fig. 
\ref{fig:seconEngFlux} and also can be seen at balloon altitude as shown in 
\cite{sark20}.
 
For the current purpose, we consider the fluxes of proton, neutron, $\gamma$, 
$e^{-}$, $e^{+}$, $\mu^{-}$, $\mu^{+}$, $\pi^{-}$ and $\pi^{+}$ generated from 
primary GCR interactions with atmosphere. In this simulation, we do not 
consider the weakly interacting (e.g. neutrinos) and low abundant ($K^{-}$, 
$K^{+}$, $\overline{p}$, $\overline{n}$ and others) particles, we also neglect 
$\pi^{-}$, $\pi^{+}$ for upward and $\pi^{o}$ for both directions due to their 
low flux. The atmospheric radiation calculations were done at 10 km altitude 
(which is the typical cruise altitude of commercial aircraft, e.g., FL328) and 
in $\sim$45$^\circ$--50$^\circ$ geomagnetic-latitude ($\theta_{M}$) region, since most 
frequent flight paths pass through this region \citep{icao16}.

\begin{figure}
\centering
\includegraphics[width=1.0\textwidth]{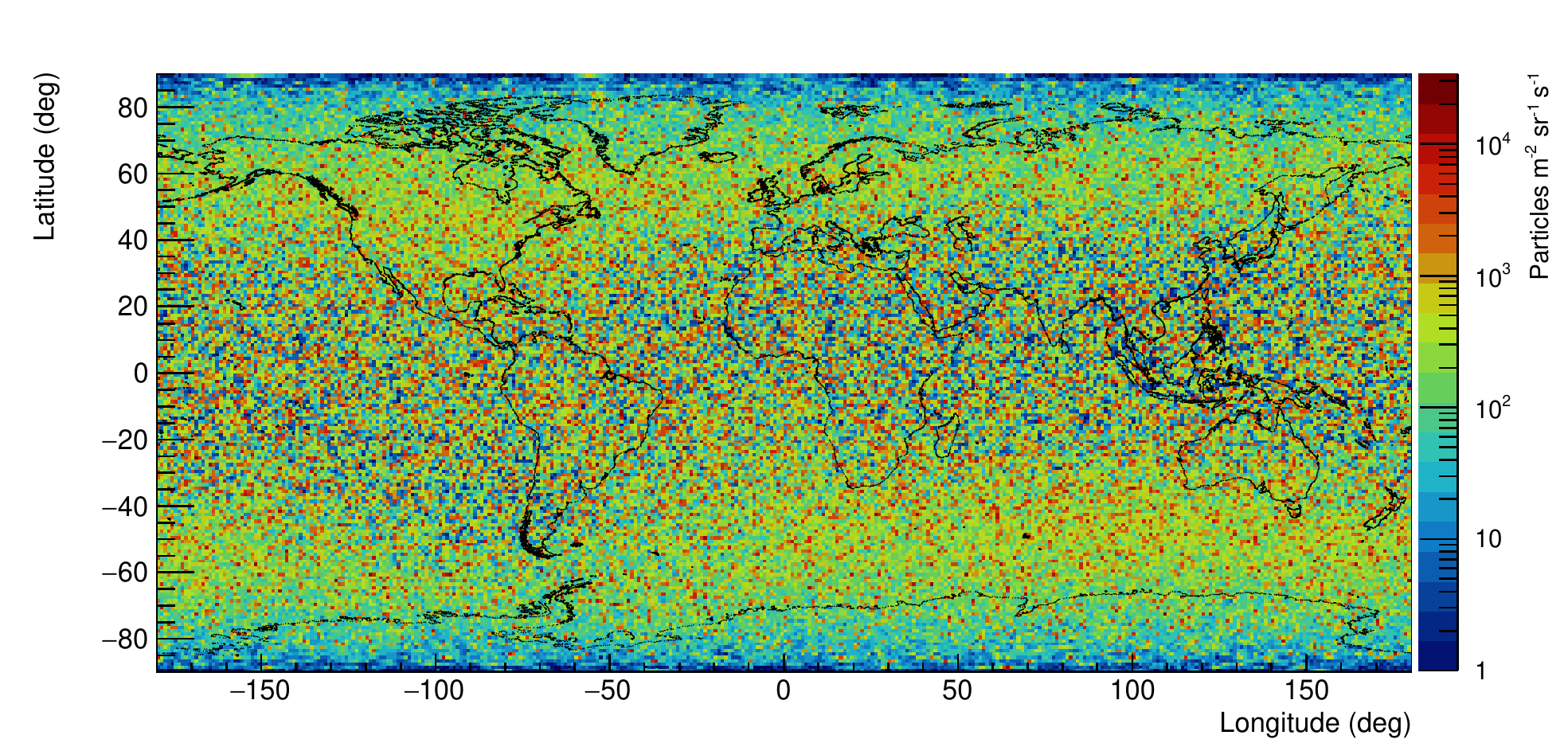}
\caption{Positional distribution of all the simulated particles and radiation in
the atmosphere at 10 km altitude over the whole geographic region.}
\label{fig:particleDist}
\end{figure}

The dose calculation at other regions and altitudes can be done in the same way
and will be considered in future. However, the latitudinal and longitudinal
distribution of all the simulated atmospheric particles at 10 km altitude can be
seen from Fig. \ref{fig:particleDist} over the whole geographic region.

\subsection{Dose calculation in human body}
\label{ssec:dose}
Here in this study, we are to calculate the effective dose ($E$), equivalent
dose ($H_{T}$) and absorbed dose ($D_{R,T}$) rates in human body from radiation
interactions at aviation altitude, where $T$ refers to particular organ or
tissue type and $R$ refers to type of radiation. The effective dose is
calculated according to the equation:
\begin{equation}
E = \sum_{T} w_{T} H_{T} = \sum_{T} w_{T} \sum_{R} w_{R} D_{R,T},
\label{Eqn:doseEqn}
\end{equation}
where $w_{T}$ is the weighting factor for organ $T$ ($\sum_{T} w_{T} = 1$), 
$w_{R}$ is radiation weighting factor and $D_{R,T}$ is absorbed dose averaged 
over the organ $T$ for particular radiation type $R$. The absorbed dose 
$D_{R,T}$ is the energy deposited per unit mass. 

The ICRP first introduced the set of $w_{R}$ values in \cite{icrp91}, which was 
updated by a new value for proton and introducing more accurate energy 
dependent continuous function of $w_{R}$ instead of a step function for the 
neutrons in \cite{icrp07}. But the consideration of constant $w_{R}$ values 
for proton and other CR ions may not be correct at higher altitudes where the 
energy and flux of these particles are higher \citep{banj19}.

\begin{figure}
\centering
\begin{subfigure} [b] {0.49\linewidth}
\includegraphics[scale=0.38]{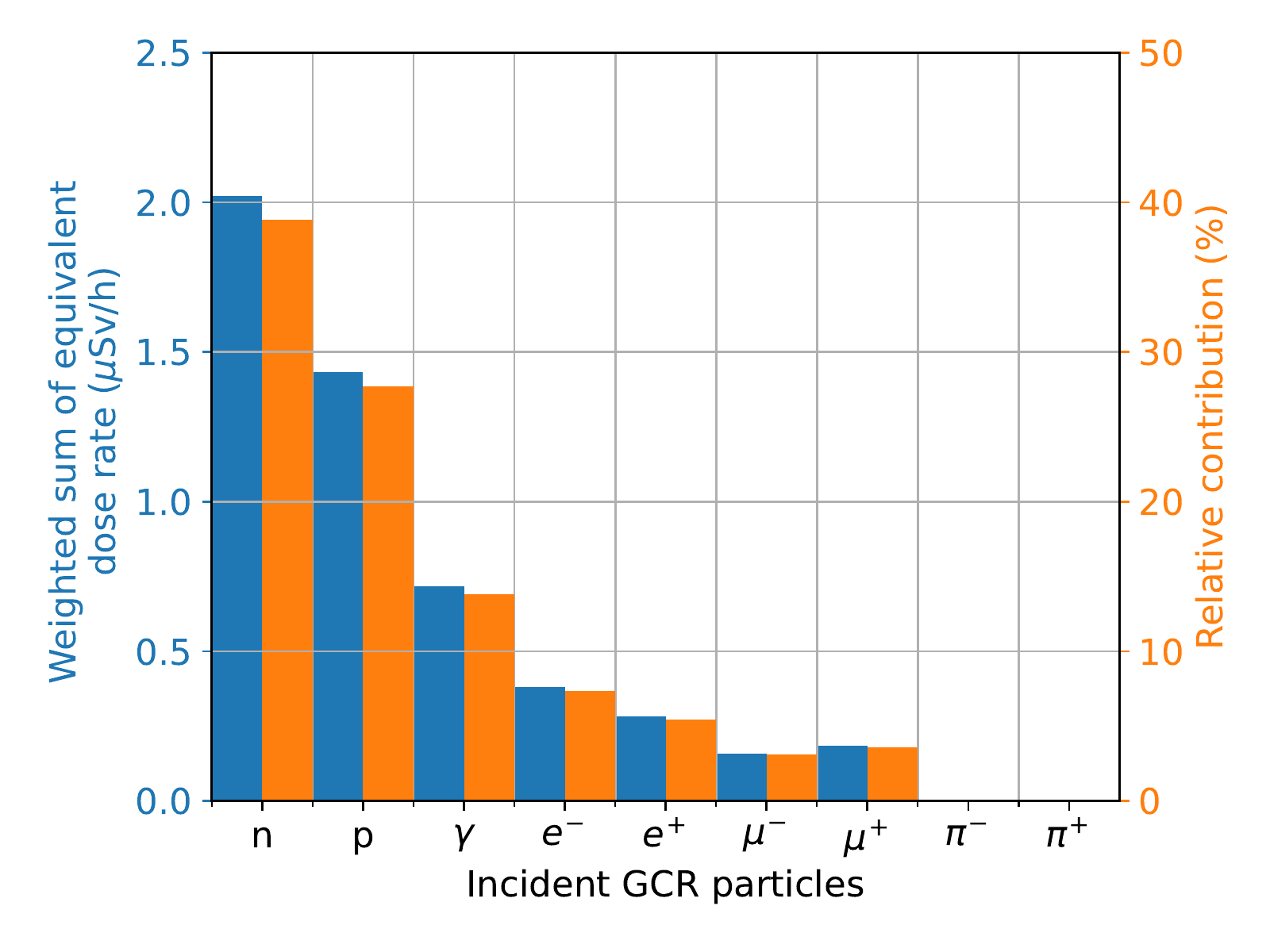}
\caption{Male}
\end{subfigure}
\begin{subfigure} [b] {0.49\linewidth}
\includegraphics[scale=0.38]{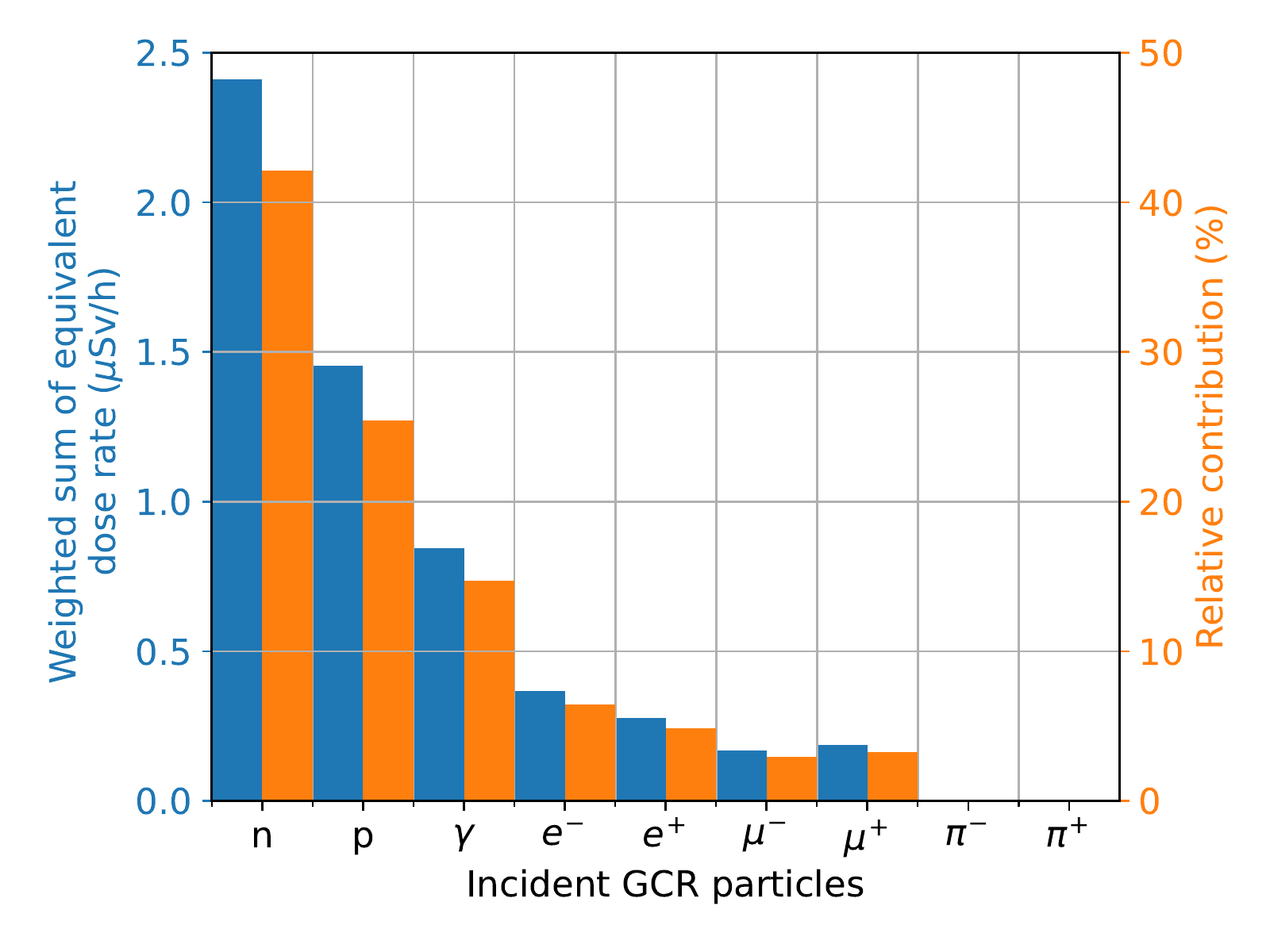}
\caption{Female}
\end{subfigure}
\caption{Weighted sum of equivalent dose rates and their relative contributions in male and
female phantoms due to different incident secondary CR particles and
radiation.}
\label{fig:primDose}
\end{figure}

We calculated the total effective dose, as well as the individual contributions
from different omnidirectional radiation and particles of atmospheric CRs
mentioned in Sec. \ref{ssec:fltalt}. The individual contributions from different
radiation components to the weighted sum of equivalent dose rate 
is plotted in Fig. \ref{fig:primDose} for both male and female phantom along 
with their relative 
contributions to the total weighted sum of equivalent dose rate. 
The same values of weighted sum of equivalent 
dose rates along with their relative contributions are tabulated in Table 
\ref{Tab:DoseT}, separated for downward and upward components of incident 
radiations. This calculation shows that most of the weighted sum of equivalent dose received by 
the phantom is due to the downward going particles (more than $\sim$95\%), 
where maximum contribution comes from neutrons followed by proton, $\gamma$, 
$e^{\pm}$, $\mu^{\pm}$ and $\pi^{\pm}$.

\cite{icrp07} defined the sex-averaged effective dose for reference person as: 
\begin{equation}
E_{ref} = \sum_{T} w_{T} \left(\frac{H^{M}_{T}+H^{F}_{T}}{2}\right),
\label{Eqn:doseEqnRef}
\end{equation}
$H^{M}_{T}$ and $H^{F}_{T}$ being the equivalent dose in different organs of 
male and female body respectively. Since, the $w_{T}$ factor for both male and 
female body in similar organs are considered same, Eqn. \ref{Eqn:doseEqnRef} 
can be rewritten as:
\begin{equation}
% E_{ref} = \sum_{T} \left(w_{T} \frac{H^{M}_{T}}{2}\right) + \sum_{T} \left(w_{T} \frac{H^{F}_{T}}{2}\right)
E_{ref}= \frac{\sum_{T} w_{T} H^{M}_{T} + \sum_{T} w_{T} H^{F}_{T}}{2},
\label{Eqn:doseEqnRefMod}
\end{equation}
which is the average of weighted sum of equivalent dose of reference male and female. 
So, for the sake of comparison, we first calculated the total weighted sum of equivalent
dose rates in reference female and male phantoms and then took the average to 
calculate the sex-averaged total effective dose rate.
 
\begin{table}
\centering
\caption{Weighted sum of equivalent dose rates [and their relative contributions] for different
incident particles and generated secondary particles in human (male/female)
phantom for both downward and upward going particles.}

\begin{small}
\begin{tabular}{C{1.5cm}C{2.1cm}C{2.1cm}C{2.1cm}C{2.1cm}}
\hline
\multirow{2}{*}{Particles} & \multicolumn{2}{c}{Male ($\mu Sv/h$) [\%]} &
\multicolumn{2}{c}{Female ($\mu Sv/h$) [\%]}\\
\cline{2-5}
& Down  & Up  & Down  & Up \\
\hline
$n$       & 1.841 [35.35] &  0.180  [3.46]  &  2.246 [39.26]  & 0.163 [2.85]\\
$p$       & 1.434 [27.53] &  0.008  [0.15]  &  1.446 [25.27]  & 0.009 [0.15]\\
$\gamma$  & 0.693 [13.32] &  0.025  [0.48]  &  0.819 [14.32]  & 0.024 [0.42]\\
$e^{-}$   & 0.377 [07.23] &  0.004  [0.09]  &  0.363 [06.35]  & 0.004 [0.08]\\
$e^{+}$   & 0.280 [05.37] &  0.003  [0.07]  &  0.274 [04.80]  & 0.003 [0.06]\\
$\mu^{-}$ & 0.159 [03.06] &  0.002  [0.05]  &  0.166 [02.90]  & 0.003 [0.05]\\
$\mu^{+}$ & 0.185 [03.56] &  0.002  [0.04]  &  0.184 [03.23]  & 0.003 [0.05]\\
$\pi^{-}$ & 0.003 [00.07] &  0.000  [0.00]  &  0.004 [00.06]  & 0.000 [0.00]\\
$\pi^{+}$ & 0.004 [00.08] &  0.000  [0.00]  &  0.005 [00.08]  & 0.000 [0.00]\\
\hline                       
\end{tabular}
\end{small}
\label{Tab:DoseT}
\end{table}

\begin{figure}
\begin{subfigure} [c] {0.24\linewidth}
\includegraphics[scale=0.25]{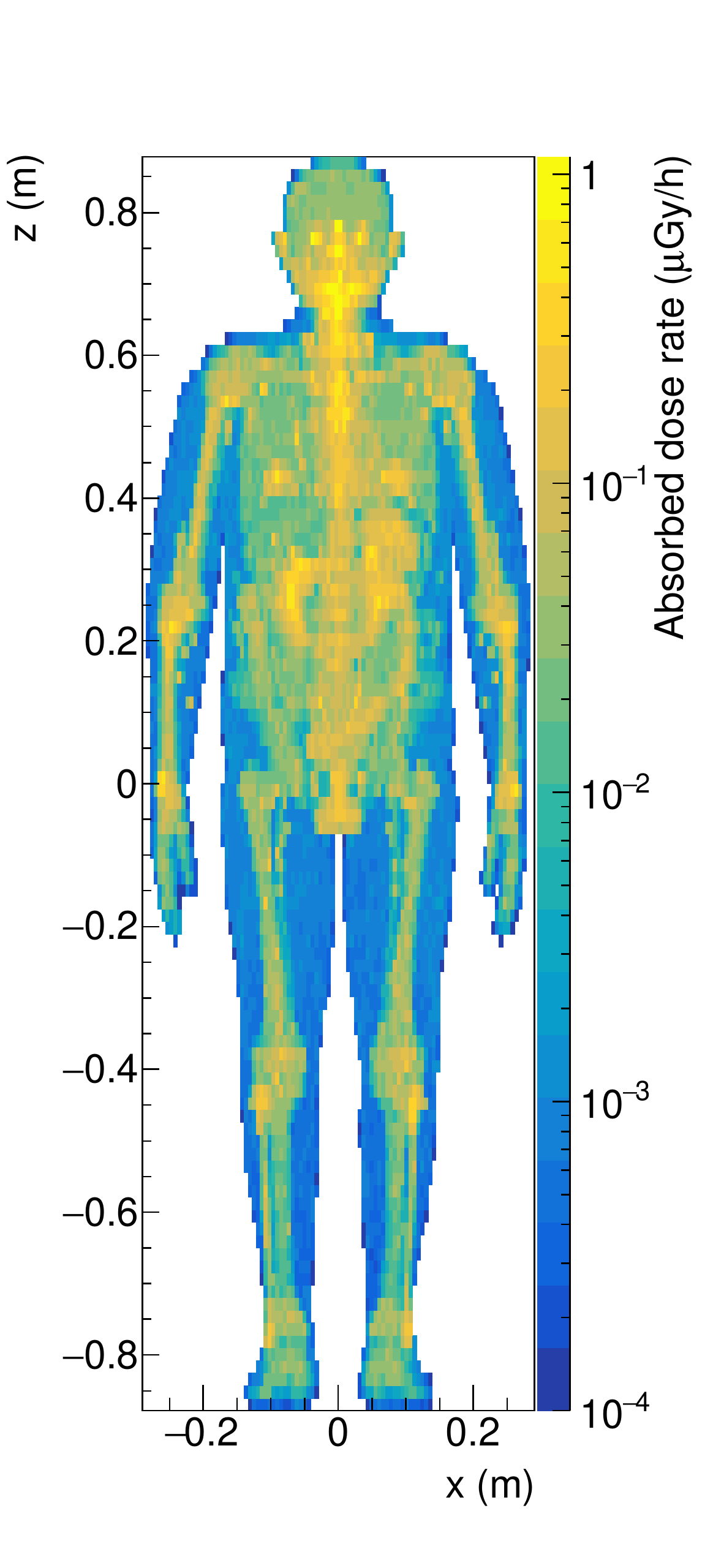}
\end{subfigure}
\begin{subfigure} [c] {0.24\linewidth}
\includegraphics[scale=0.25]{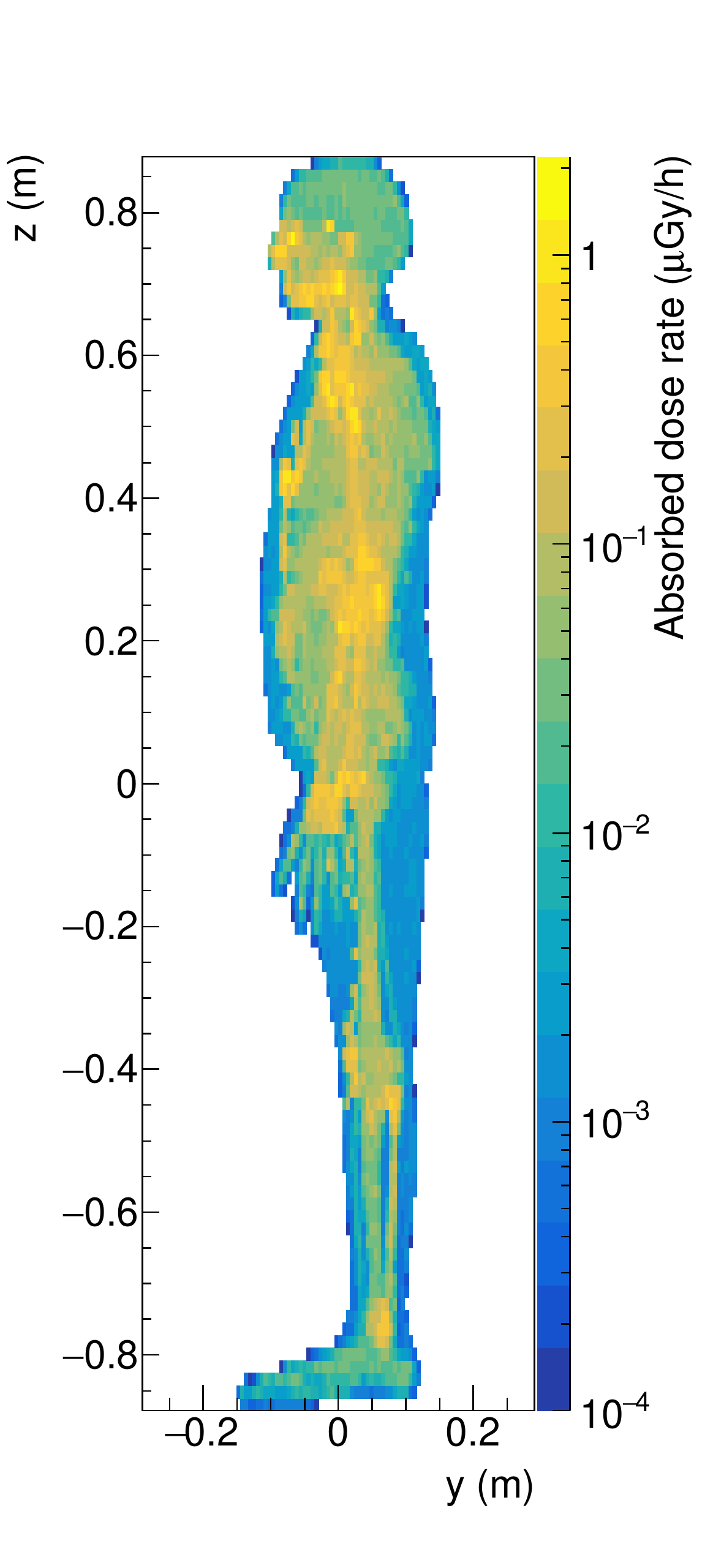}
\end{subfigure}
\begin{subfigure} [c] {0.24\linewidth}
\includegraphics[scale=0.25]{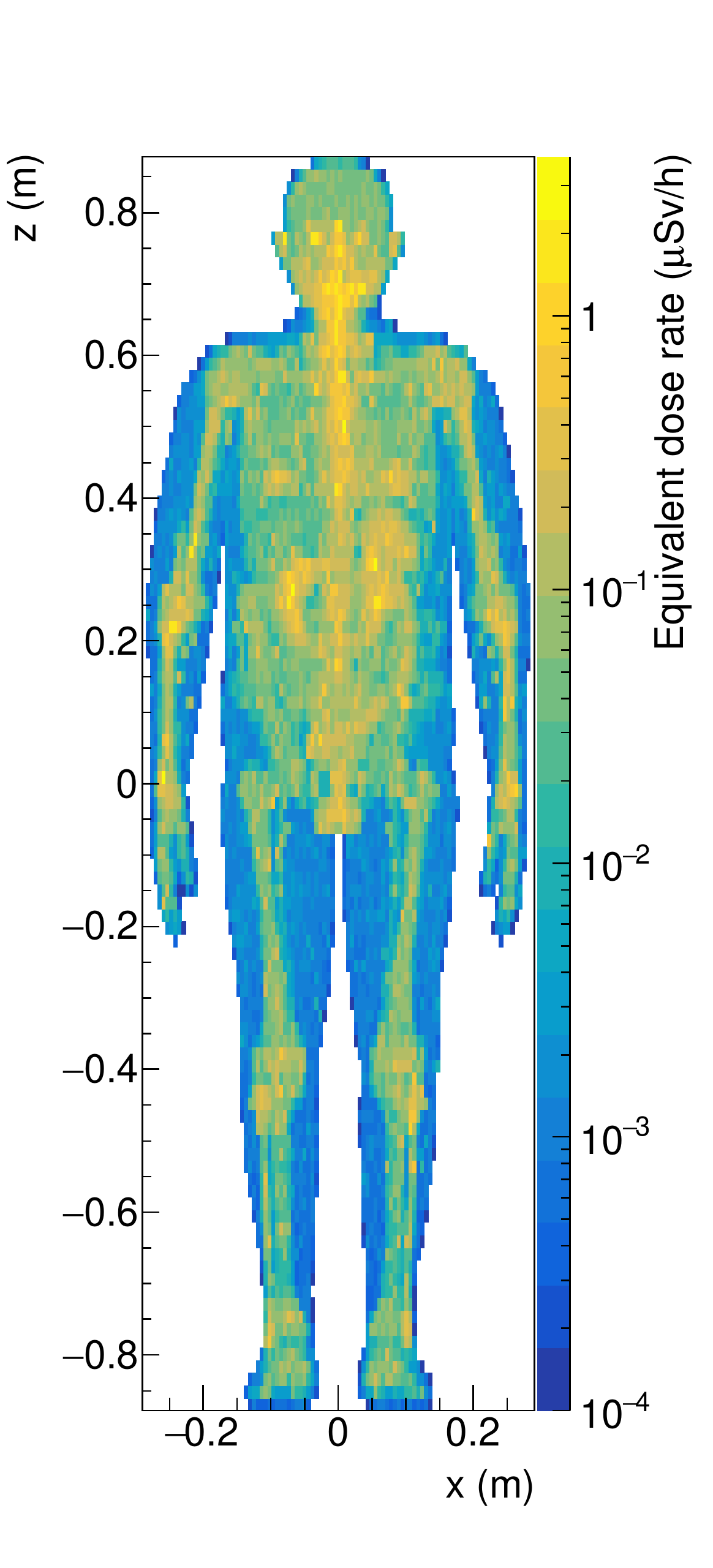}
\end{subfigure}
\begin{subfigure} [c] {0.24\linewidth}
\includegraphics[scale=0.25]{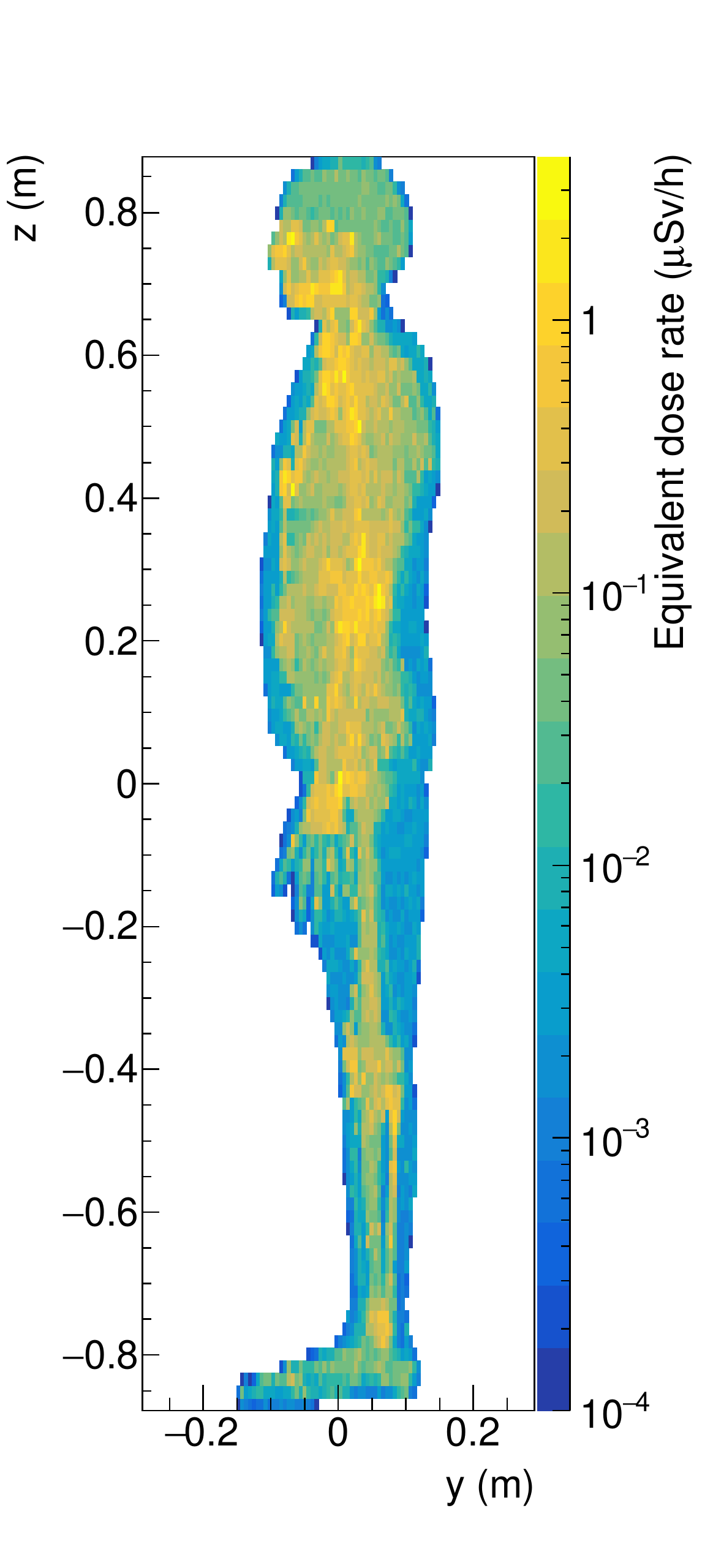}
\end{subfigure}
\caption{Spatial distribution (frontal view; side-wise view) of
absorbed dose rate and equivalent dose rate in male phantom.}
\label{fig:spMale}
\end{figure}

\begin{figure}
\begin{subfigure} [c] {0.24\linewidth}
\includegraphics[scale=0.23]{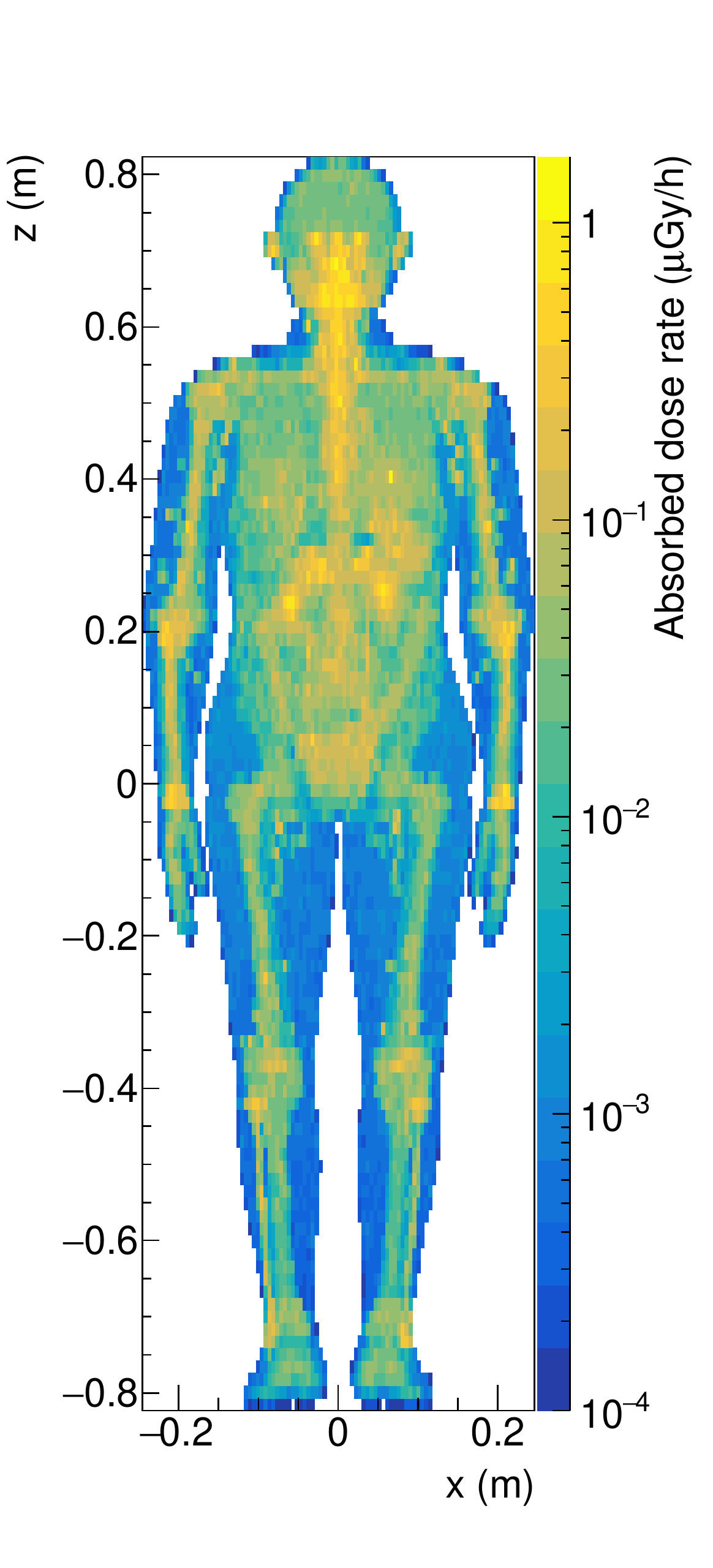}
\end{subfigure}
\begin{subfigure} [c] {0.24\linewidth}
\includegraphics[scale=0.23]{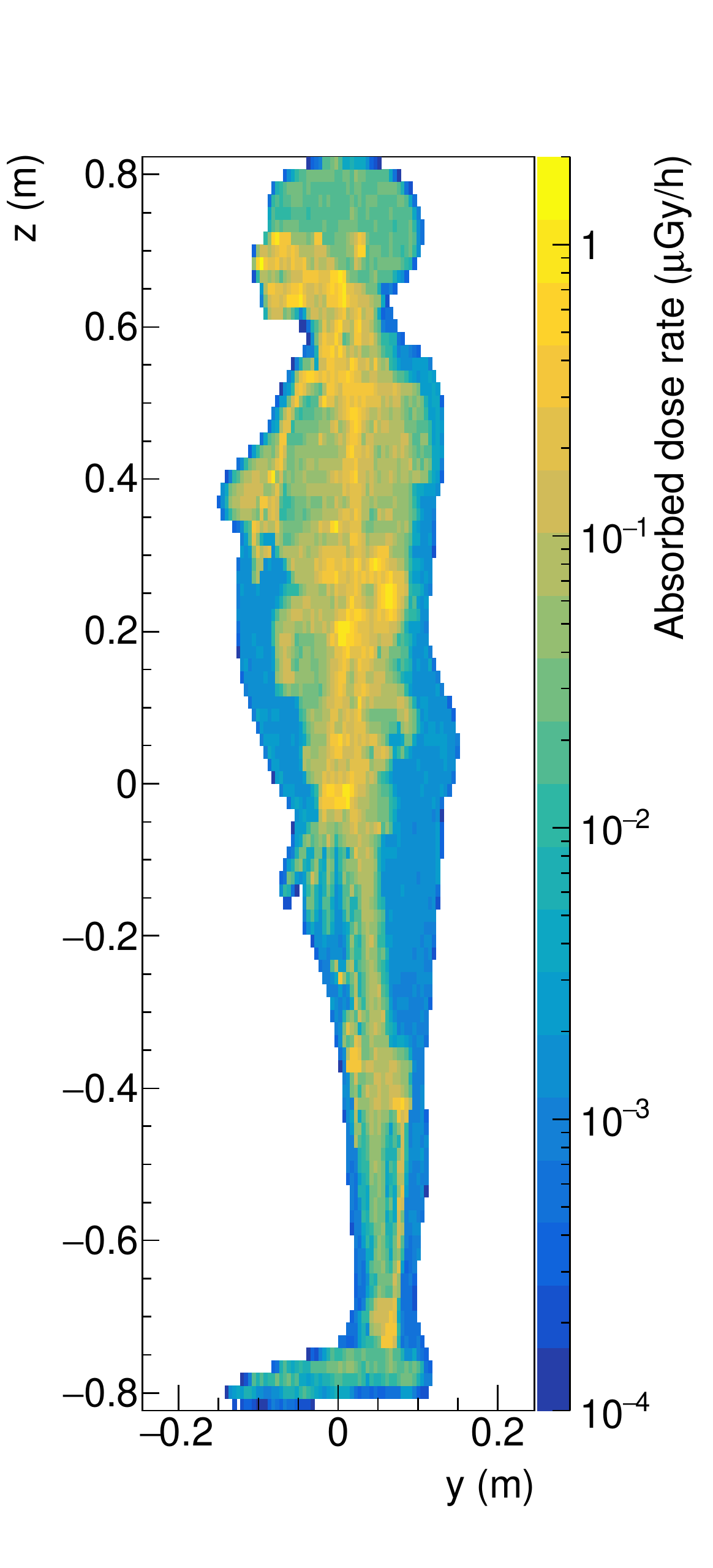}
\end{subfigure}
\begin{subfigure} [c] {0.24\linewidth}
\includegraphics[scale=0.23]{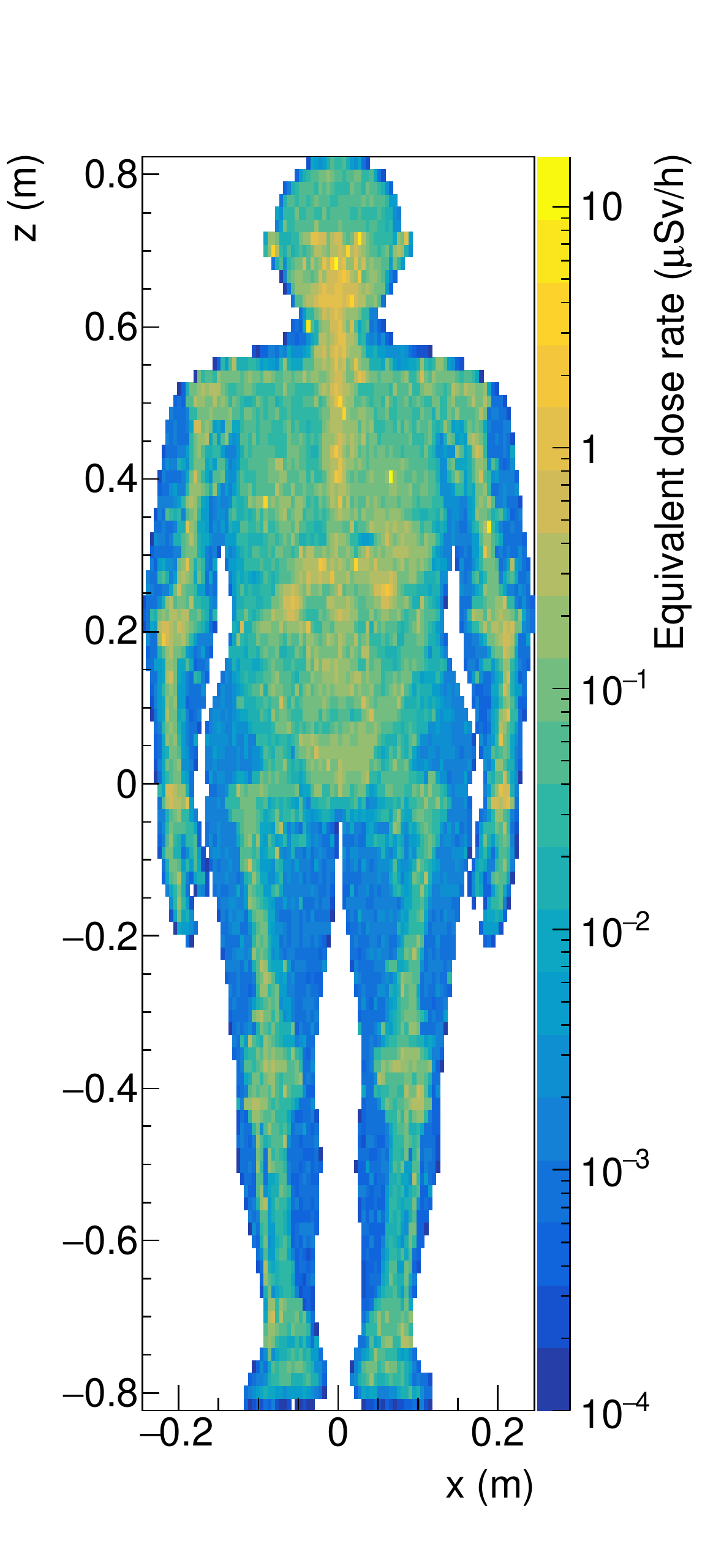}
\end{subfigure}
\begin{subfigure} [c] {0.24\linewidth}
\includegraphics[scale=0.23]{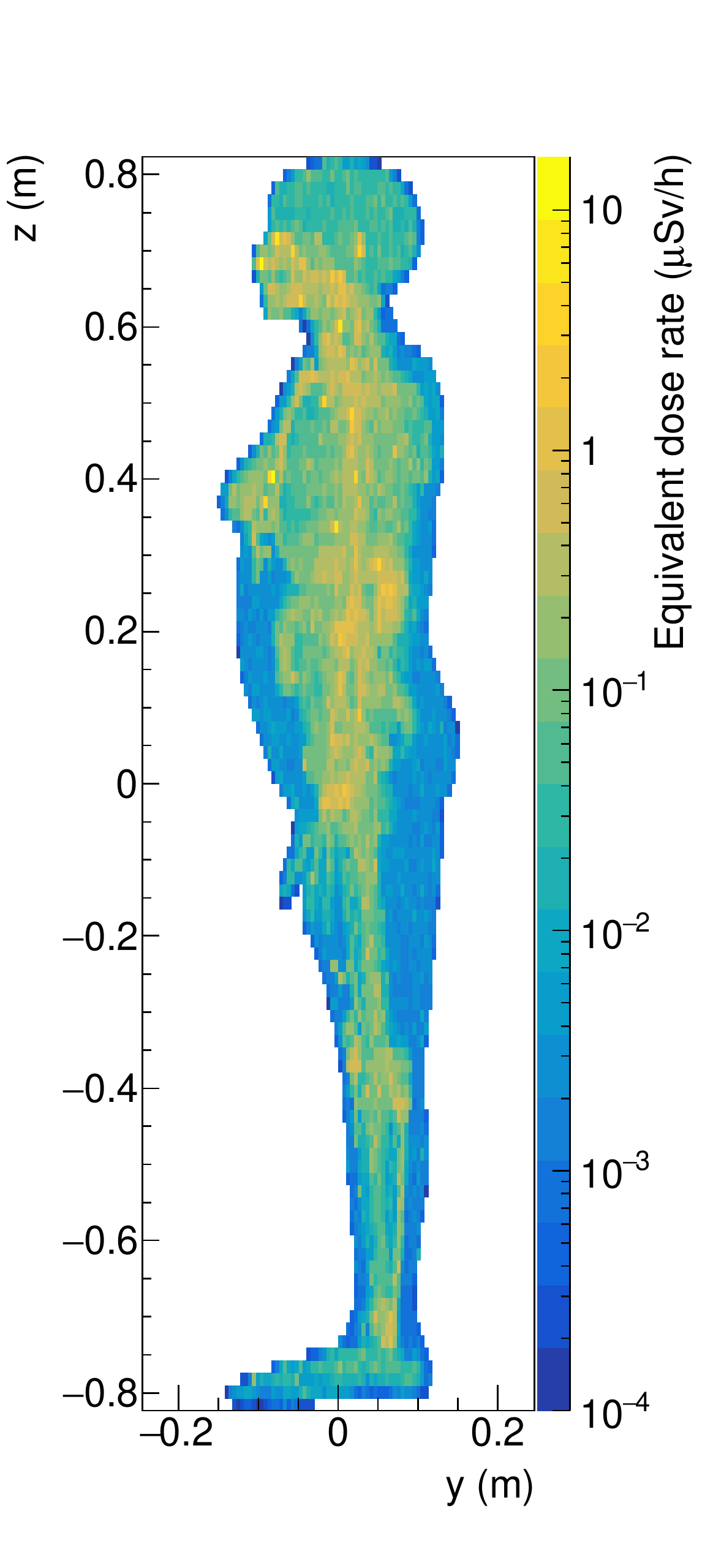}
\end{subfigure}
\caption{Spatial distribution (frontal view; side-wise view) of
absorbed dose rate and equivalent dose rate in female phantom.}
\label{fig:spFemale}
\end{figure}

\begin{figure}
\includegraphics[scale=0.38]{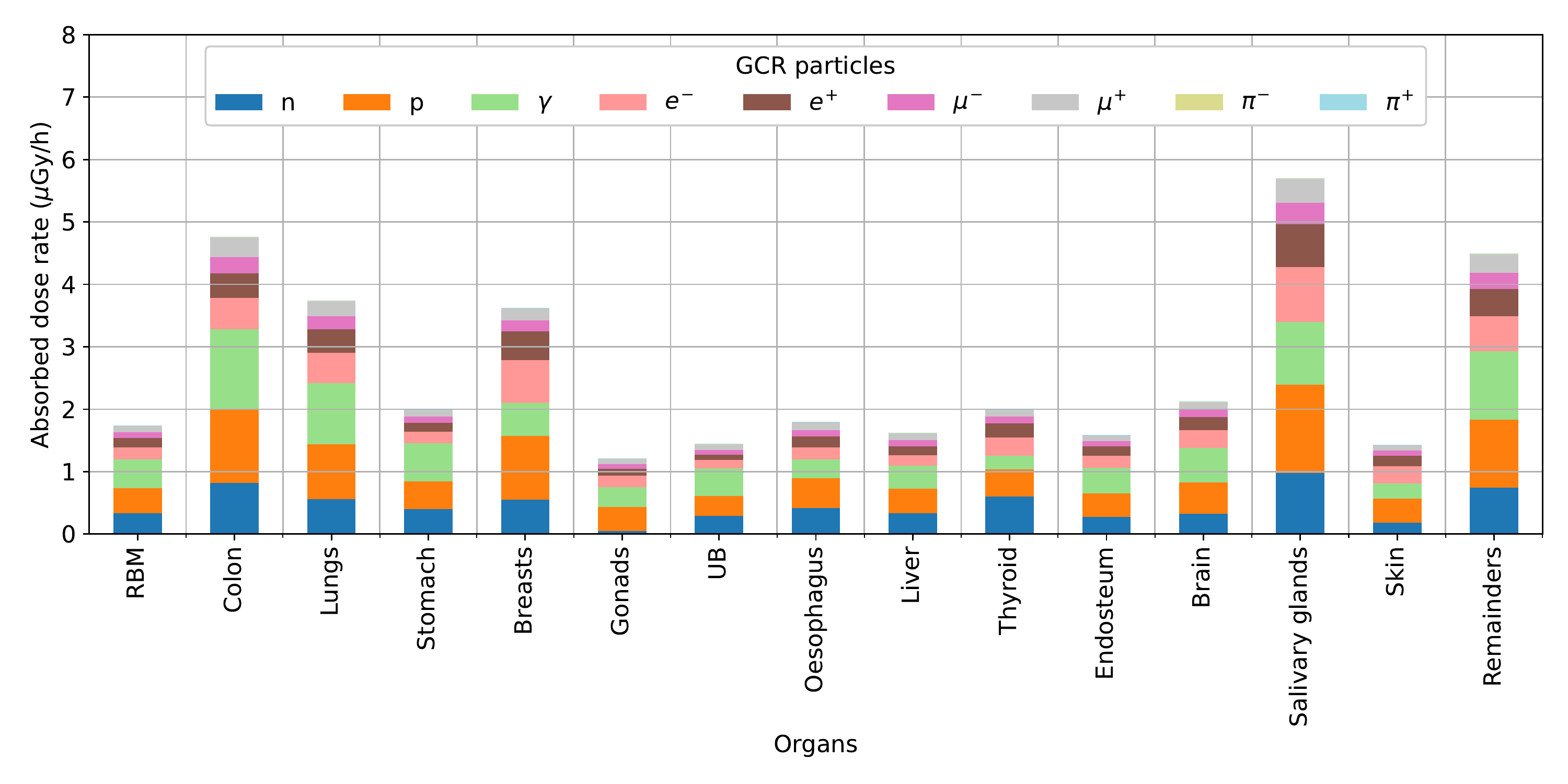}
\caption{Absorbed dose rate for different incident particles in different organs
of male phantom.}
\label{fig:absorOrgDoseM}
\end{figure}

\begin{figure}
\includegraphics[scale=0.38]{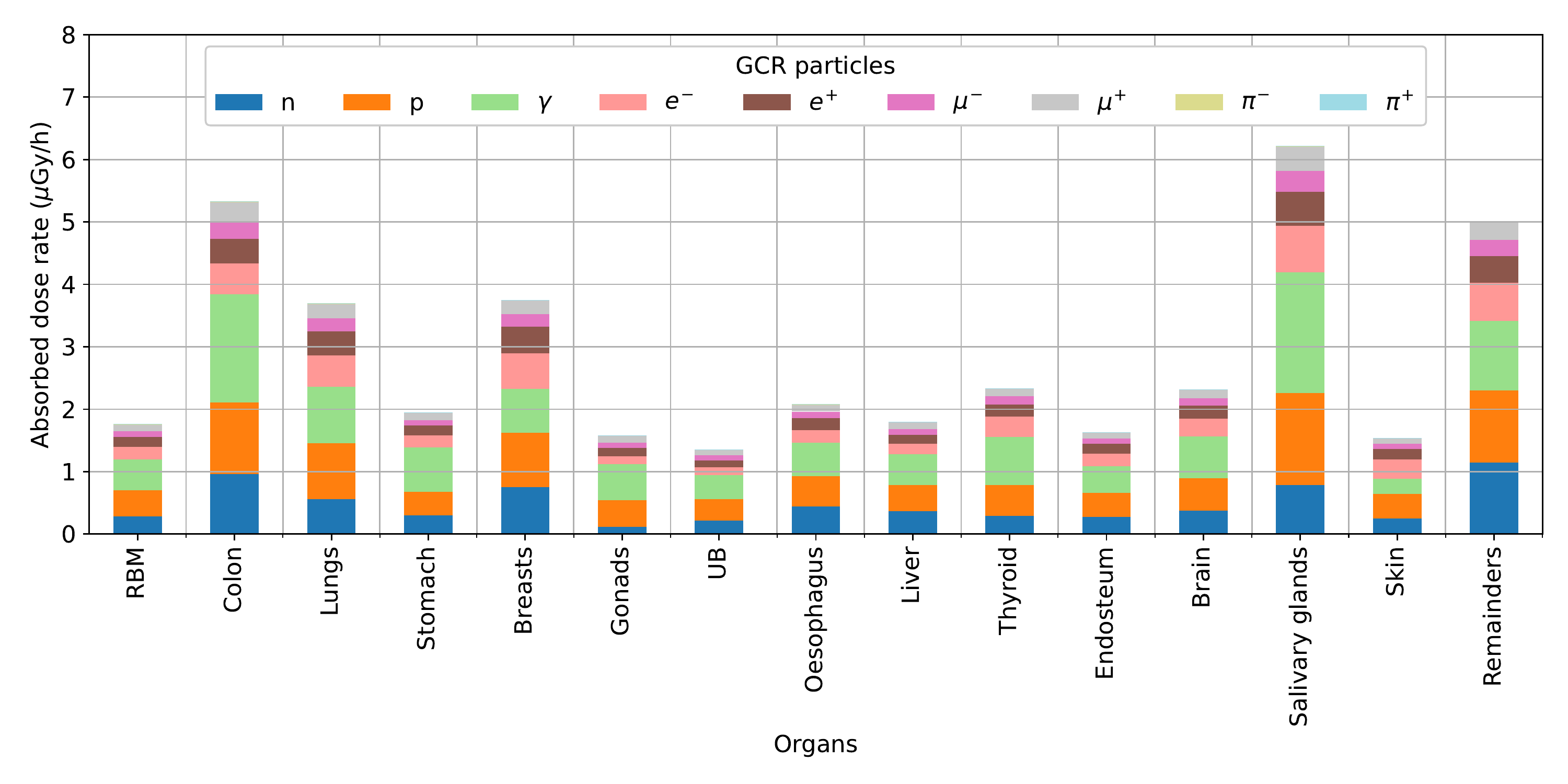}
\caption{Absorbed dose rate for different incident particles in different organs
of female phantom.}
\label{fig:absorOrgDoseF}
\end{figure}

\begin{figure}
\includegraphics[scale=0.38]{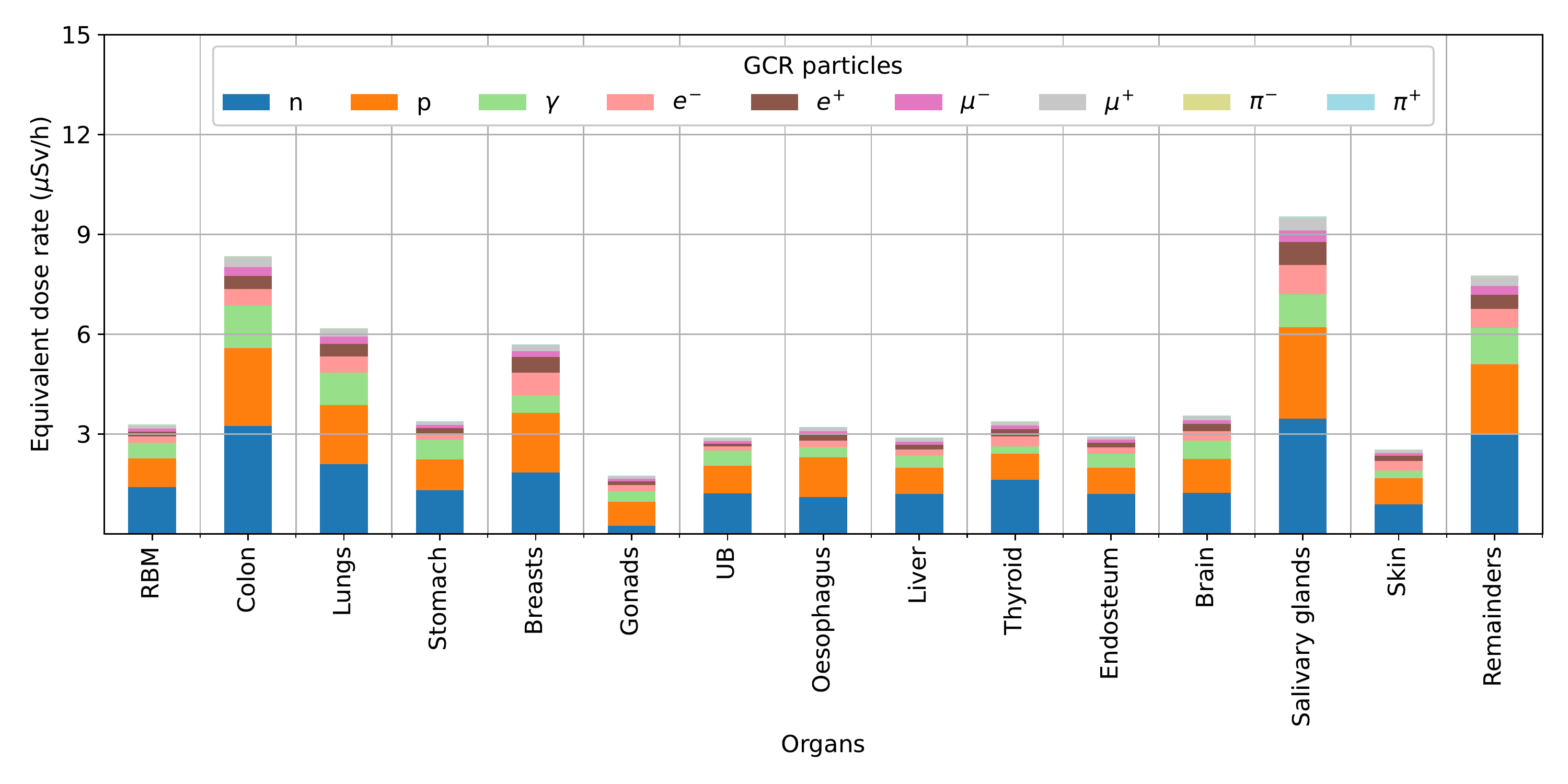}
\caption{Equivalent dose rate for different incident particles in different
organs of male phantom.}
\label{fig:equiOrgDoseM}
\end{figure}

\begin{figure}
\includegraphics[scale=0.38]{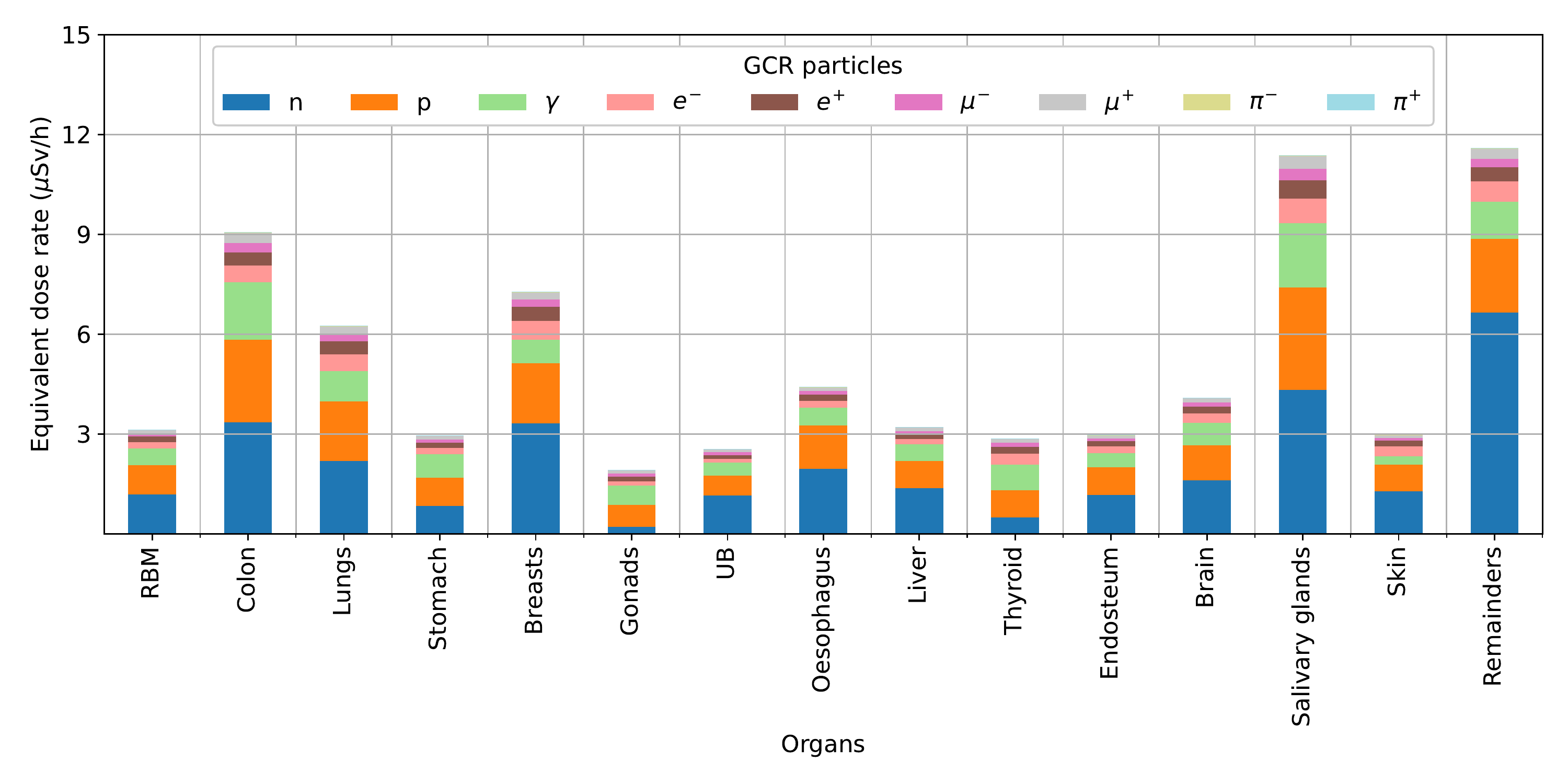}
\caption{Equivalent dose rate for different incident particles in different
organs of female phantom.}
\label{fig:equiOrgDoseF}
\end{figure}

Spatial distribution (frontal and side-wise view) of absorbed dose rate and
equivalent dose rate inside both male and female phantoms are plotted in Fig.
\ref{fig:spMale} and \ref{fig:spFemale}. The contributions to absorbed dose
rates and equivalent dose rates from different primary incident particles in 
each organs, for both male and female phantom, are also shown in Fig.
\ref{fig:absorOrgDoseM}, \ref{fig:absorOrgDoseF} and \ref{fig:equiOrgDoseM},
\ref{fig:equiOrgDoseF}. Under the same irradiation condition, the dose received 
by the female phantom is slightly higher than the male phantom. Noticeable 
difference in radiation dose rates at different organs is quite apparent from 
these plots. \citet{lund19} also find similar variation in organ dose using a 
human anatomical phantom for Apollo 11 and Apollo 14 missions from trapped 
protons. Using a MC simulation procedure with simplified phantom model in X-ray 
radiation environment for medical purpose, \cite{lewi88} reported different 
absorption dose rates in different organs indicating a ratio more than factor 
of 2 between colon and stomach. In a recent study of organ dose estimation due 
to computed tomography, \cite{dema20} also reported substantial difference of 
radiation dose rates in different organs which also support our calculation. 
However, \cite{reit12} reported more or less flat distribution of absorbed 
dose rate at different organs in human body at the radiation environment on 
moon with only high energy primary GCRs and using ICRP phantom model. We can 
speculate the differential organ doses due to radiation environment at the 
aviation altitude and organ material distribution in the phantoms (high density
and relatively high Z bone material should play a role here). The direct 
verification of the obtained result is difficult due to lack of experimental 
knowledge and other calculations of dose rate for individual organs at this 
level. Some future development in this concern with computation or/and 
experiments can shed more light and verify the results obtained in the current 
calculation.

The sex-averaged absorbed (equivalent) dose rate on eye lens is calculated as
1.84 $\mu$Gy/h (5.16 $\mu$Sv/h). The equivalent dose limit of eye lens for
occupational exposure and public exposure is 20 mSv and 15 mSv per year. So 
from this simulation, it can be seen that this limit can be exceeded for 
frequent fliers or aircrews which could increase the chance of eye cataracts. 
The absorbed dose rate and equivalent dose rate received by the prostate 
(uterus) of male (female) phantom is 1.25 (1.78) $\mu$Gy/h and 1.82 (3.21) 
$\mu$Sv/h. A recent study by \citet{graj15} suggests that the risk of 
miscarriage among the female flight attendant will increase for radiation dose 
of or more than 0.1 mGy in week 9--13 from conception. The estimated threshold 
dose for the gestational period can be found in \citet{pate07}, which is quite 
high but can be reached during severe solar particle events, Terrestrial 
Gamma-ray Flash events \citep{dwye10} or frequent use of flights from high 
altitude or long haul polar routes. So, all frequently flying pregnant women 
or aircrews should take extra care of their fetus during their early gestation.

The incident energetic particles and radiation produce different sub-atomic
secondary particles, as well as some radioactive nuclei, by interacting with the
human body. These radioactive nuclei are mainly produced by spallation reaction
of protons and activation by neutrons in our body \citep{brod69}. The amount of
radiation dose per hour imparted by these radionuclides is negligible, but can
be accumulated to give substantial dose for extended exposure time or elevated
radiation environment. Because of their radioactive nature, these cosmogenic
radio-isotopes are particularly important, as they can be accumulated in some
specific organs due to their physicochemical nature, where it could cumulatively
give radiation dose depending upon their biological half-lives which may lead to
stochastic detrimental effect like cancer. Despite of the harmful characteristics 
of the cosmogenic radionuclides, they could be used to accurately measure the 
radiation dose in the human body \citep{brod69}. They have half-lives ranging 
from less than a second to years. Internal radioactivity is particularly 
dangerous than the external radiation as the decaying particles or radiations 
may kill nearby tissue, cells or damage the DNA structure. In the current 
simulation work we also have calculated the production of radioactive nuclei 
in human body due to high energy particles or radiation interactions. Most 
abundant nuclides among all the produced radionuclides in human body are listed 
in Table \ref{tab:cosmo} along with their half-life, mode of production and 
abundance.

\begin{table}
\centering
\caption{Total production of cosmogenic radionuclides per hour inside phantoms
from interaction of cosmic ray at aircraft altitude.}
\begin{tabular}{C{0.18\textwidth}C{0.20\textwidth}C{0.28\textwidth}C{0.20\textwidth}}
\hline
Radionuclide & Half-life  & Mode of production$^{*}$ & Atoms/hour\\% & Decay mechanism 
\hline
$^{14}$C	 & 5700 y               & $^{14}N(n,p)^{14}C$              & 2.88$\times10^{5}$ \\ %& $\beta^{-}$     
$^{15}$O	 & 122.24 s             & $^{16}O(n,2n)^{15}O$             & 1.24$\times10^{5}$ \\ %& ec $\beta^{+}$  
$^{36}$Cl  & 3.01$\times10^{5}$ y & $^{35}Cl(n,\gamma)^{36}Cl$       & 8.39$\times10^{4}$ \\ %& $\beta^{-}$     
$^{11}$C	 & 20.36 m              & $^{12}C(n,2n)^{11}C$             & 6.59$\times10^{4}$ \\ %& ec $\beta^{+}$  
$^{7}$Be	 & 53.22 d              & $^{12}C(n,2n+\alpha)^{7}Be$      & 3.20$\times10^{4}$ \\ %& ec              
$^{40}$K	 & 1.24$\times10^{9}$ y & $^{39}K(n,\gamma)^{40}K$         & 1.91$\times10^{4}$ \\ %& $\beta^{-}$     
$^{14}$O	 & 12.5 ms              & $^{16}O(n,3n)^{14}O$             & 2.08$\times10^{4}$ \\ %& $\beta^{+} $    
$^{6}$He	 & 806.7 ms             & $^{12}C(n,p+\alpha+d)^{6}He$     & 1.45$\times10^{4}$ \\ %& $\beta^{-}$     
$^{16}$N	 & 7.13 s               & $^{16}O(n,p)^{16}N$              & 1.16$\times10^{4}$ \\ %& $\beta^{-}$     
$^{41}$Ca	 & 9.94$\times10^{4}$ y & $^{40}Ca(n,\gamma)^{41}Ca$       & 1.25$\times10^{4}$ \\ %& ec              
$^{10}$Be	 & 1.51$\times10^{6}$ y & $^{12}C(n,2n+p)^{10}B$           & 1.27$\times10^{4}$ \\ %& $\beta^{-}$     
$^{12}$B	 & 20.20 ms             & $^{16}O(n,p+\alpha)^{12}B$       & 6.75$\times10^{3}$ \\ %& $\beta^{-}$     
$^{13}$N	 & 9.96 m               & $^{16}O(n,p+3n)^{13}N$           & 1.09$\times10^{4}$ \\ %& ec $\beta^{+}$  
$^{8}$Li	 & 839.9 ms             & $^{12}C(n,p+\alpha)^{8}Li$       & 8.20$\times10^{3}$ \\ %& $\beta^{-}$     
$^{37}$Ar  & 35.01 d              & $^{40}Ca(n,\alpha)^{37}Ar$       & 6.97$\times10^{3}$ \\ %& ec              
$^{12}$N	 & 11.0 ms              & $^{16}O(n,p+4n)^{12}N$           & 5.75$\times10^{3}$ \\ %& ec $\beta^{+}$  
$^{32}$P	 & 14.26 d              & $^{31}P(n,\gamma)^{32}P$         & 6.04$\times10^{3}$ \\ %& $\beta^{-}$     
$^{24}$Na	 & 14.99 h              & $^{23}Na(n,\gamma)^{24}Na$       & 4.03$\times10^{3}$ \\ %& $\beta^{+}$     
\hline
\end{tabular}
\\
\footnotesize{$^{*}$ Only major production processes}
\label{tab:cosmo}
\end{table}

Based on this result, we tried to have a rough estimation of the maximum limit 
to the career exposure of the aircrews due to cosmogenic radionuclides listed 
in Table \ref{tab:cosmo}. For this purpose, we considered the decay of the 
induced radionuclides at their corresponding position of generation inside 
human body as obtained from the simulation and calculated the absorbed dose 
from different radionuclides. The total number of radionuclides accumulated 
in the body is calculated considering overall occupational exposure for 20 
and 30 years with 1000 flight hours per year. The number of radionuclides that 
decay over the lifetime after the exposure (considering 50 years from the 
median age of occupational exposure suffered) to give the radiation dose, 
depending on the corresponding half lives of the radionuclides. This 
calculation yields the carrier exposure of about 6 mSv and 9 mSv considering 
20 and 30 years of occupational exposures respectively. However, it should be 
emphasized here that, this calculation is based on the physical half-lives of 
the radionuclides. But due to various reasons their biological half-lives 
would be different, which should be actually considered in the calculation. 
Again, the production rates of these radionuclides depends on the location 
(altitude/latitude) of exposure over the occupational career which also should 
be considered.

\subsection{Radiation flux inside aircraft structure}
\label{ssec:radflux}
\begin{figure}
\begin{subfigure}[b]{1.0\textwidth}
\centering
\includegraphics[scale=0.3]{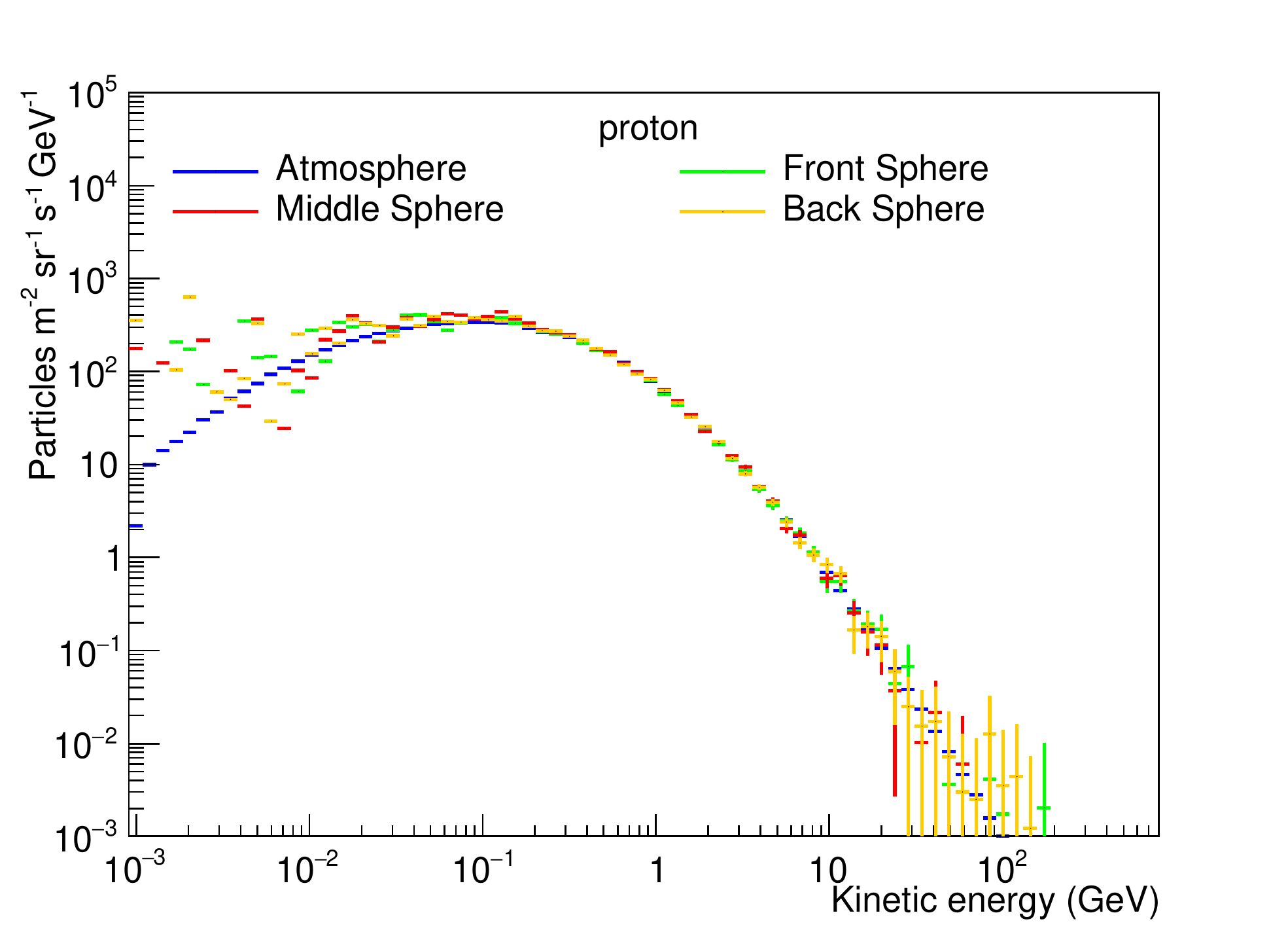}
\end{subfigure}%
\\
\begin{subfigure}[b]{0.49\textwidth}
\centering
\includegraphics[scale=0.3]{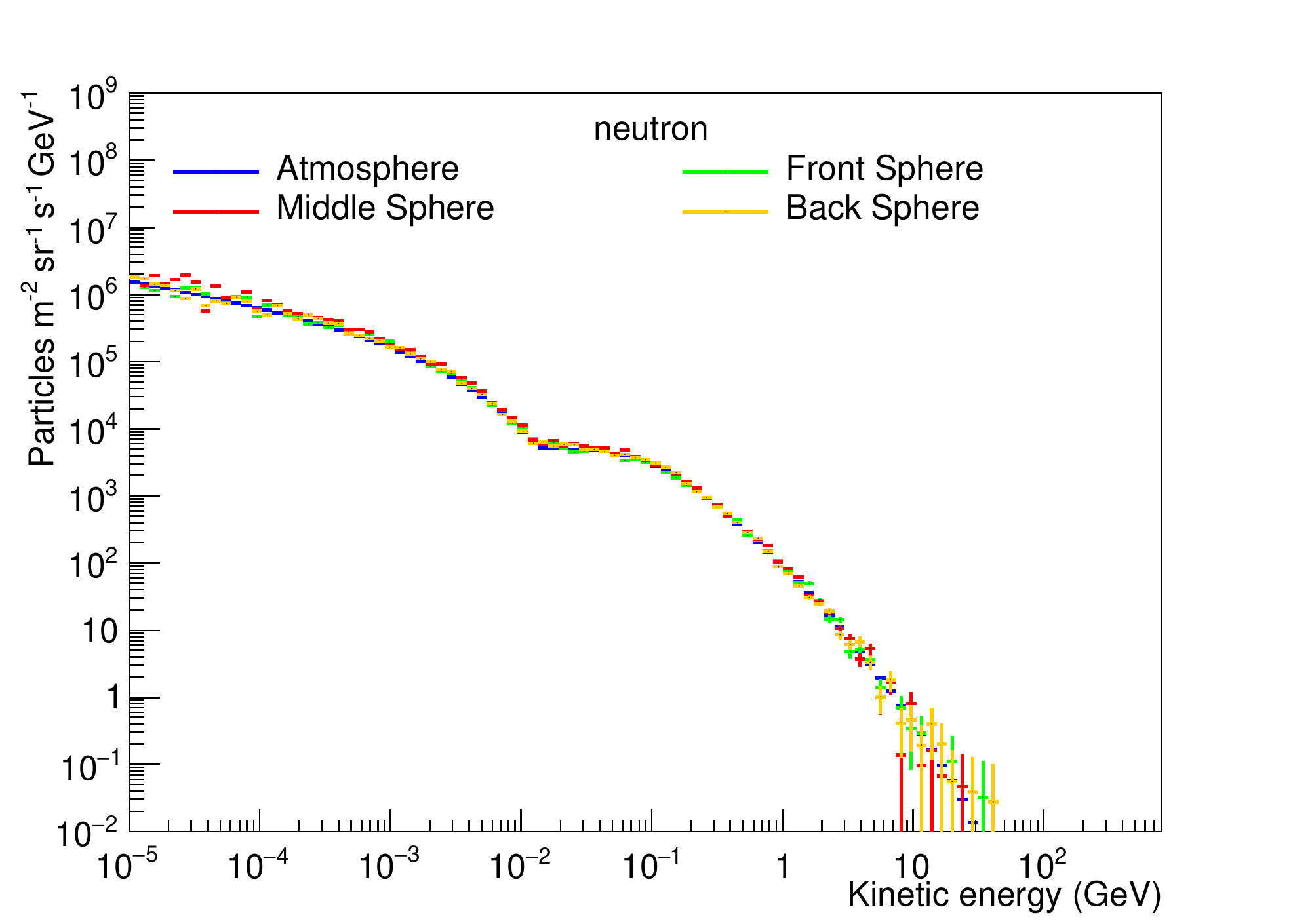}
\end{subfigure}
\begin{subfigure}[b]{0.49\textwidth}
\centering
\includegraphics[scale=0.3]{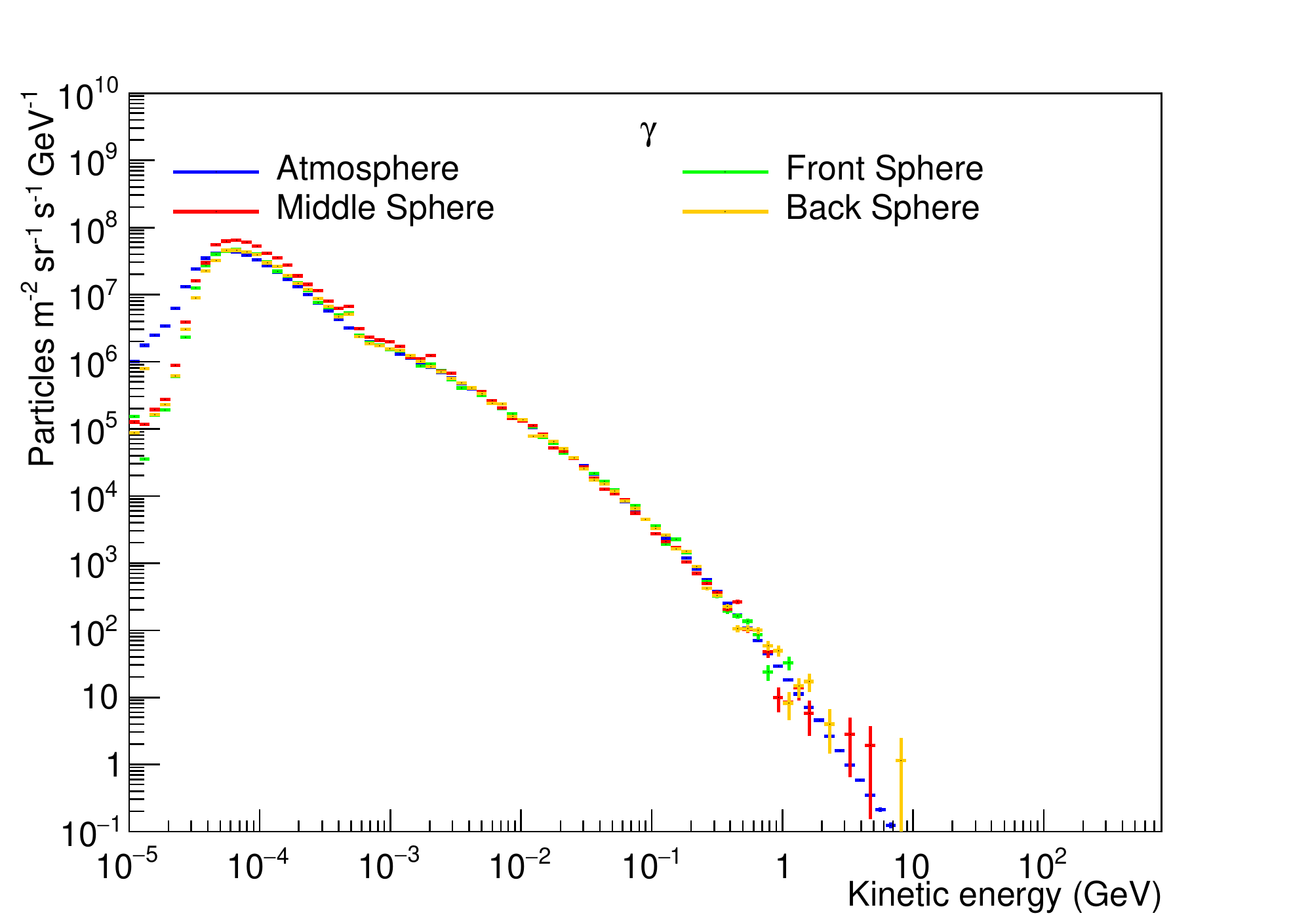}
\end{subfigure}
\begin{subfigure}[b]{0.49\textwidth}
\centering
\includegraphics[scale=0.3]{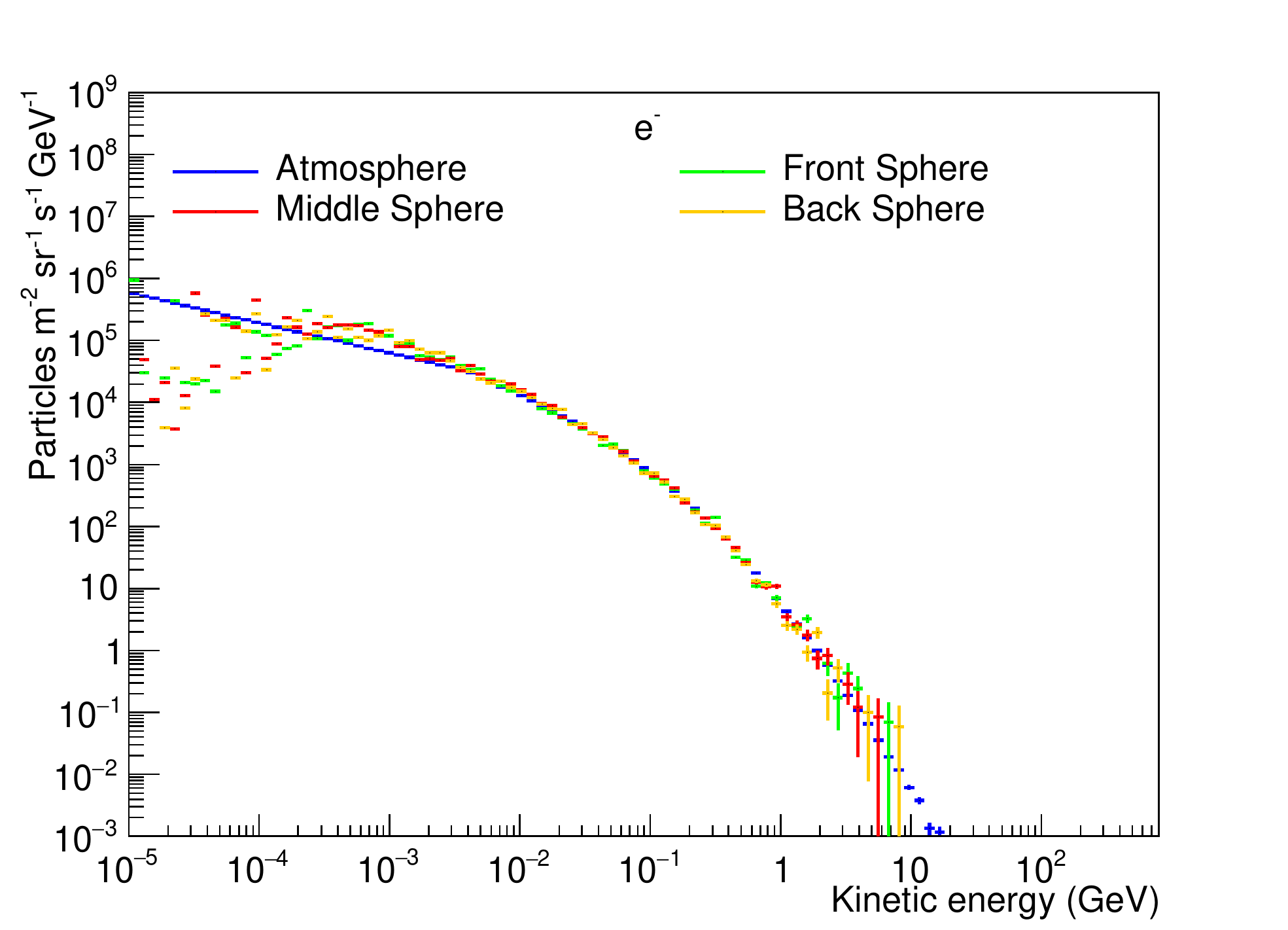}
\end{subfigure}
\begin{subfigure}[b]{0.49\textwidth}
\centering
\includegraphics[scale=0.3]{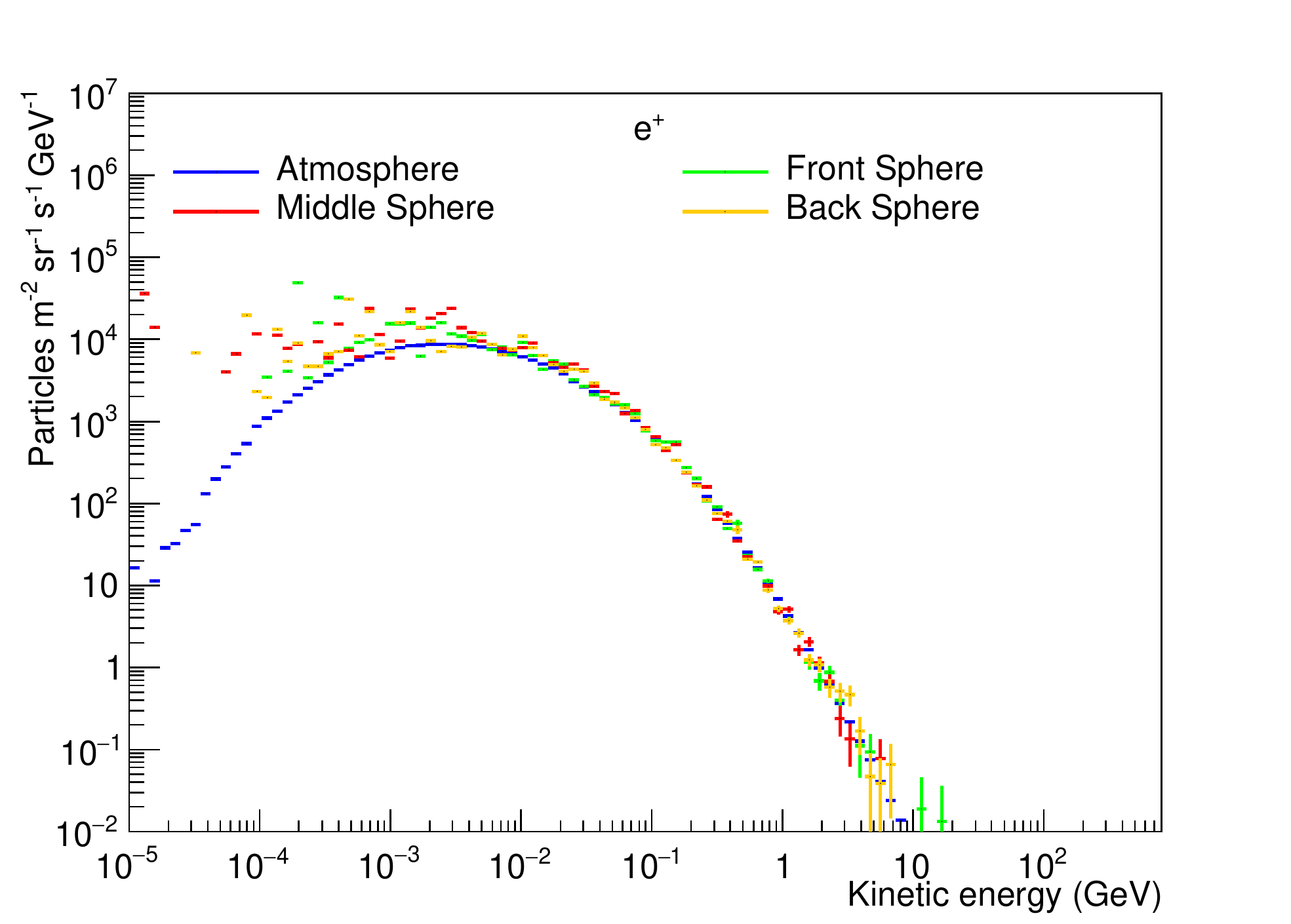}
\end{subfigure}
\begin{subfigure}[b]{0.49\textwidth}
\centering
\includegraphics[scale=0.3]{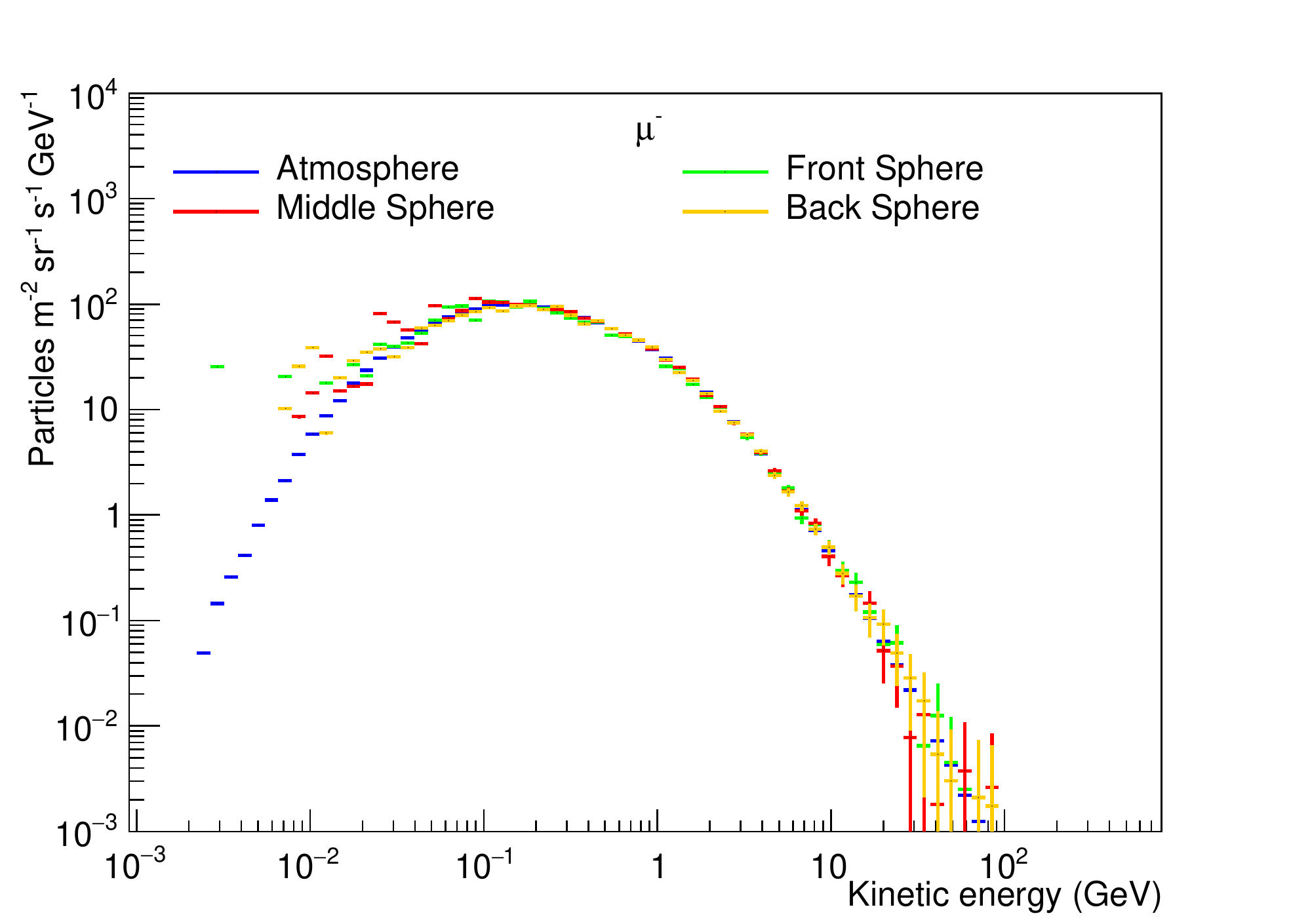}
\end{subfigure}
\begin{subfigure}[b]{0.49\textwidth}
\centering
\includegraphics[scale=0.3]{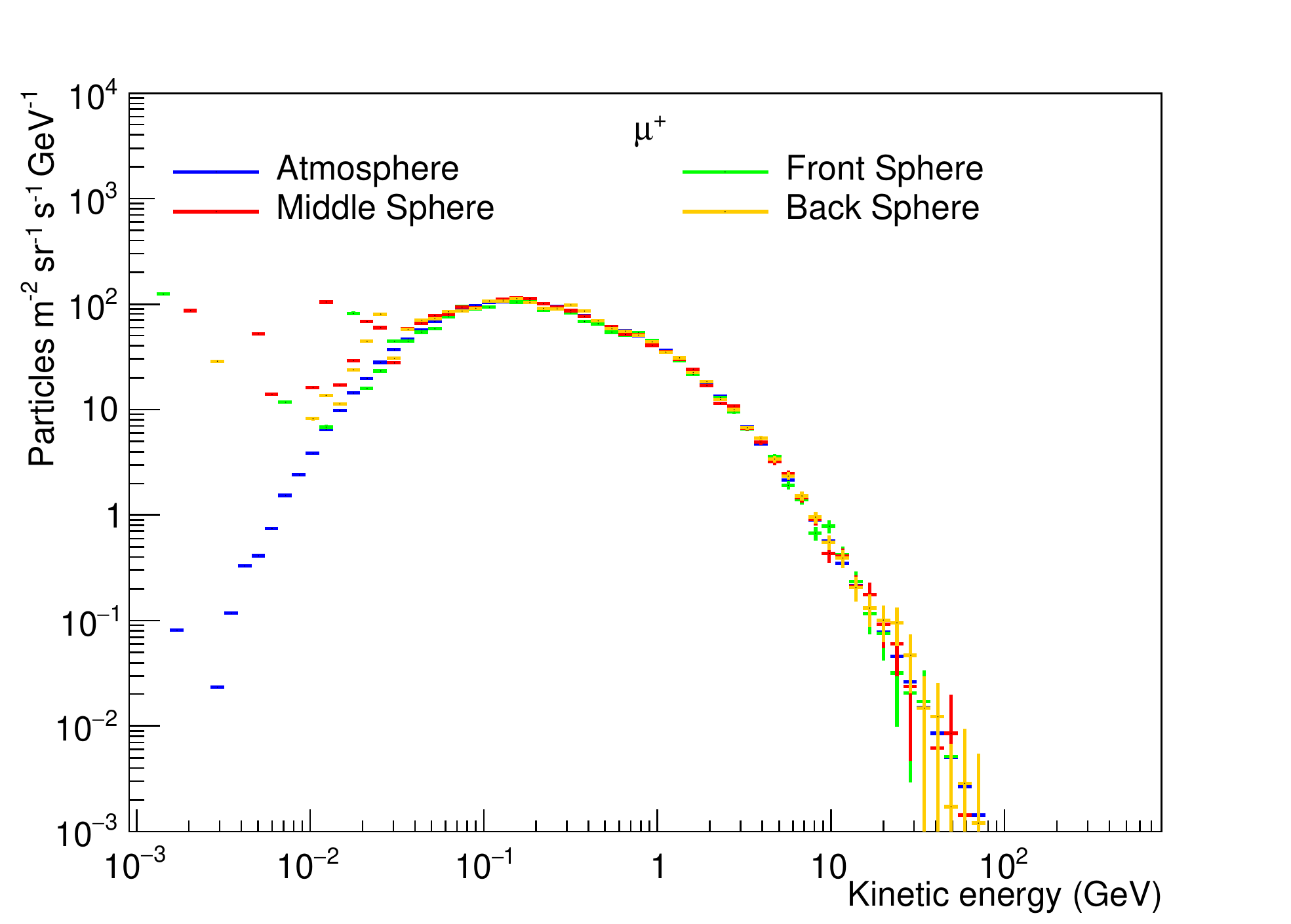}
\end{subfigure}
\caption{Particle flux for different downward going secondary GCR particles (proton, neutron,
$\gamma$, $e^{-}$, $e^{+}$, $\mu^{-}$, $\mu^{+}$) at aviation altitude and in four
different positions (outside:Atmosphere, front, middle and back side of the aircraft).}
\label{fig:radfldn}
\end{figure}

\begin{figure}
\begin{subfigure}[b]{1.0\textwidth}
\centering
\includegraphics[scale=0.3]{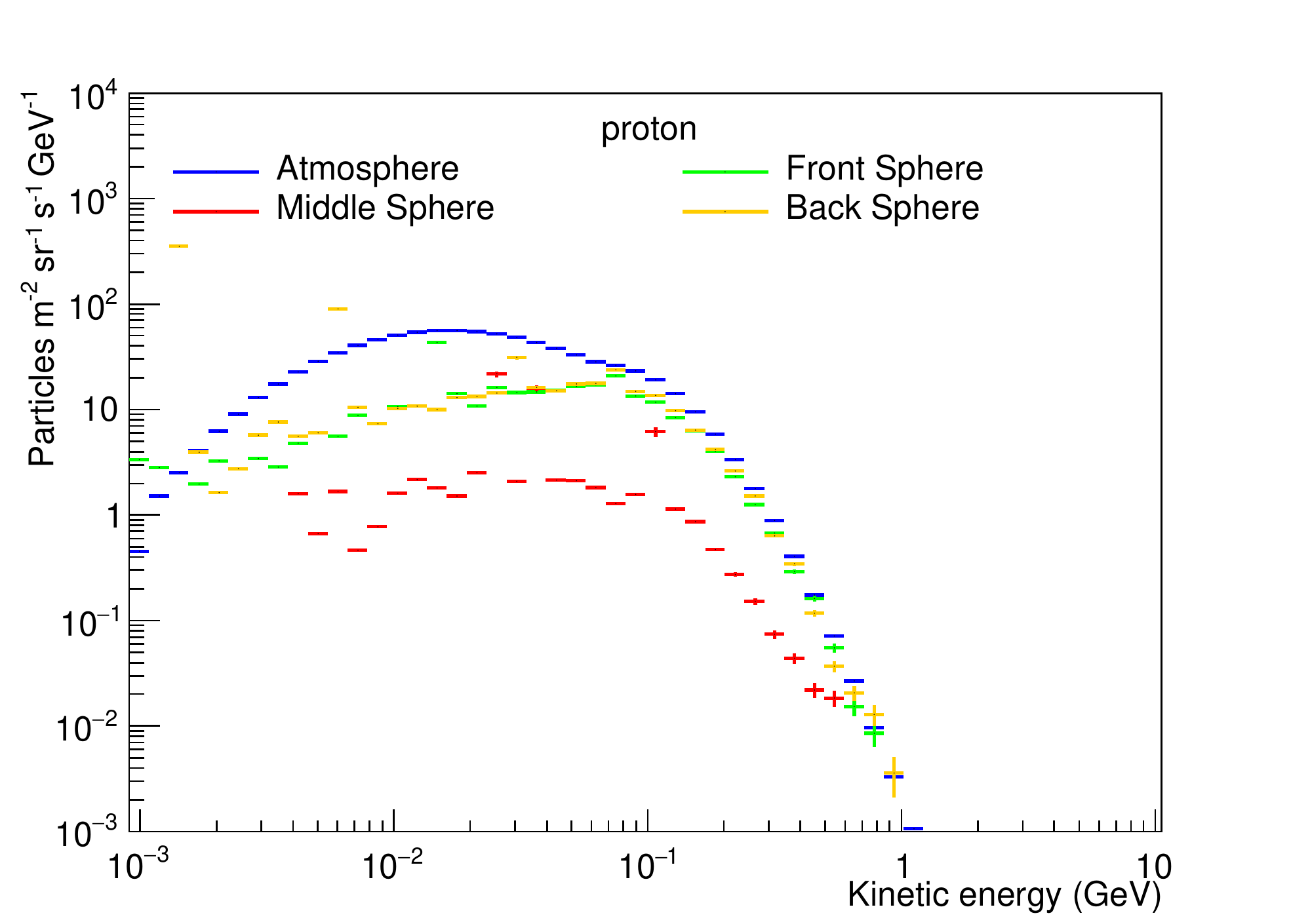}
\end{subfigure}
\\
\begin{subfigure}[b]{0.49\textwidth}
\centering
\includegraphics[scale=0.3]{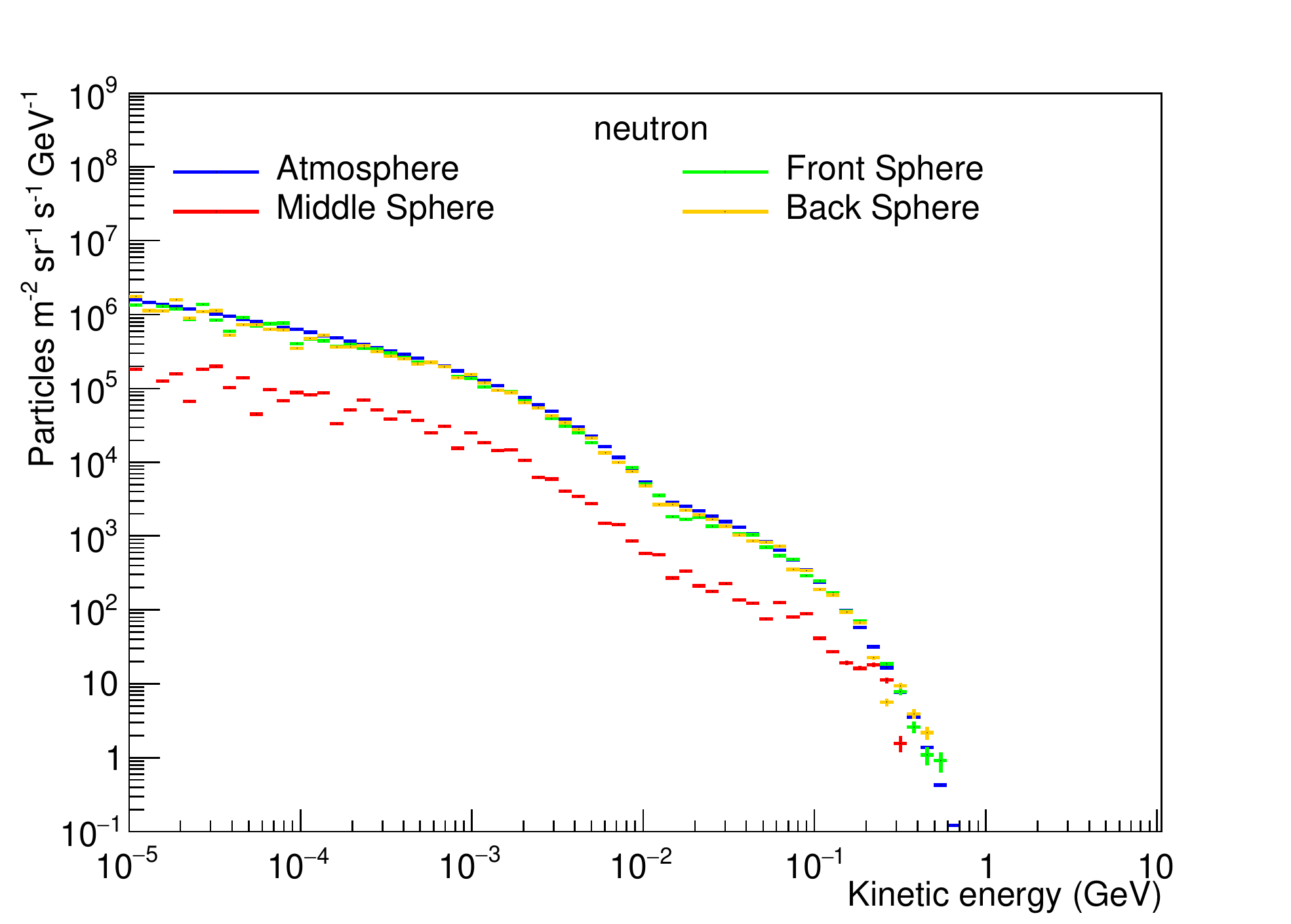}
\end{subfigure}
\begin{subfigure}[b]{0.49\textwidth}
\centering
\includegraphics[scale=0.3]{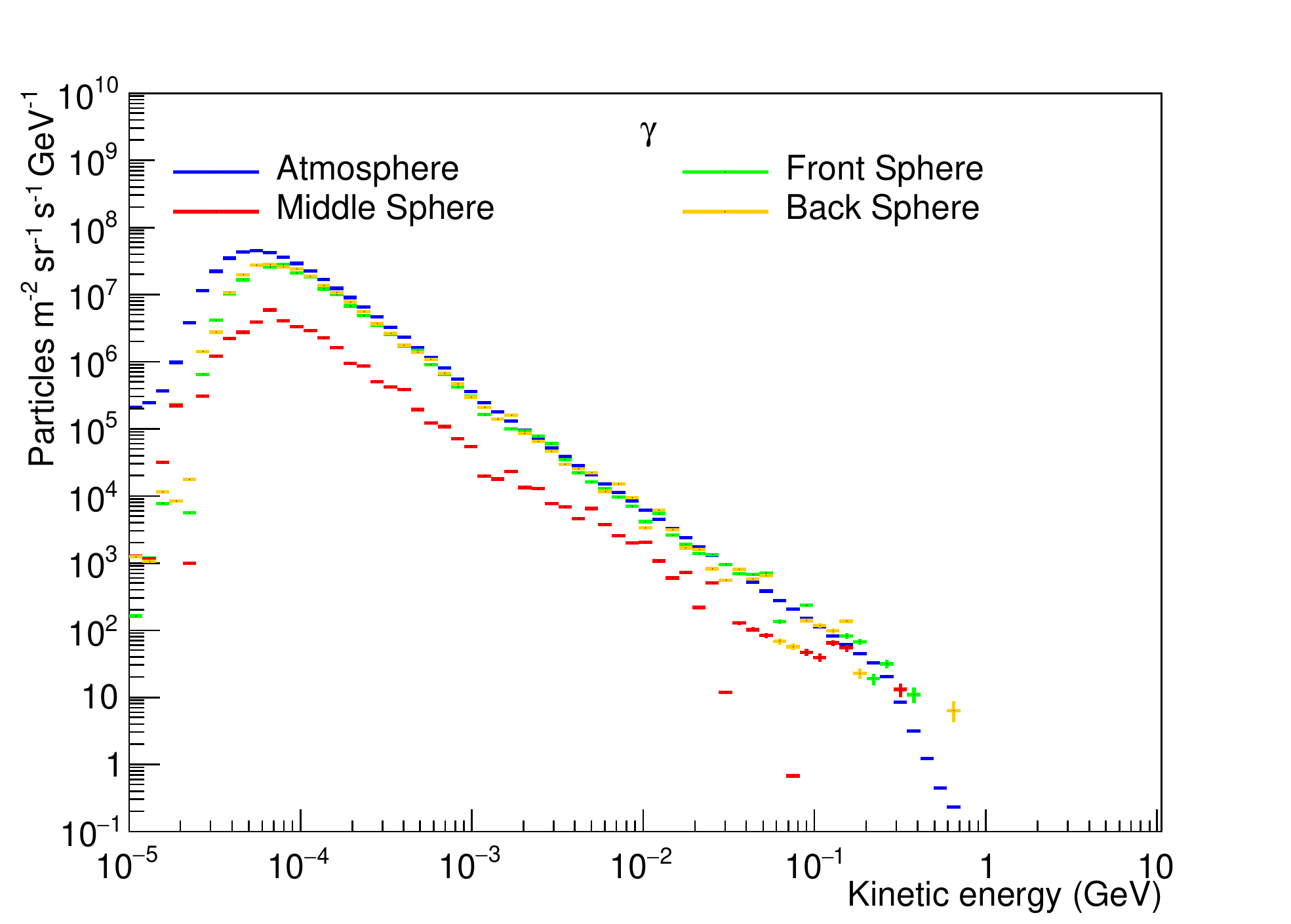}
\end{subfigure}
\begin{subfigure}[b]{0.49\textwidth}
\centering
\includegraphics[scale=0.3]{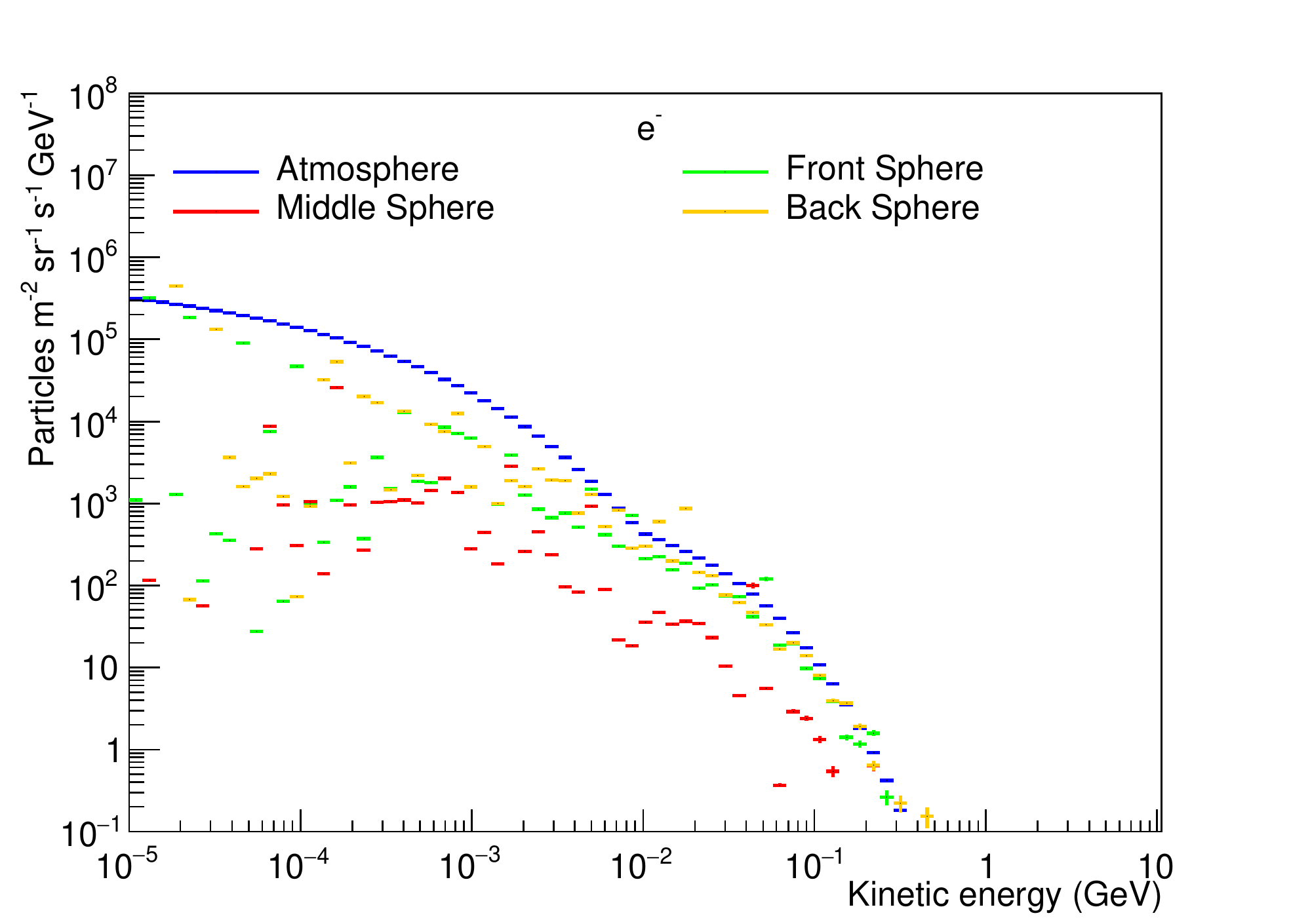}
\end{subfigure}
\begin{subfigure}[b]{0.49\textwidth}
\centering
\includegraphics[scale=0.3]{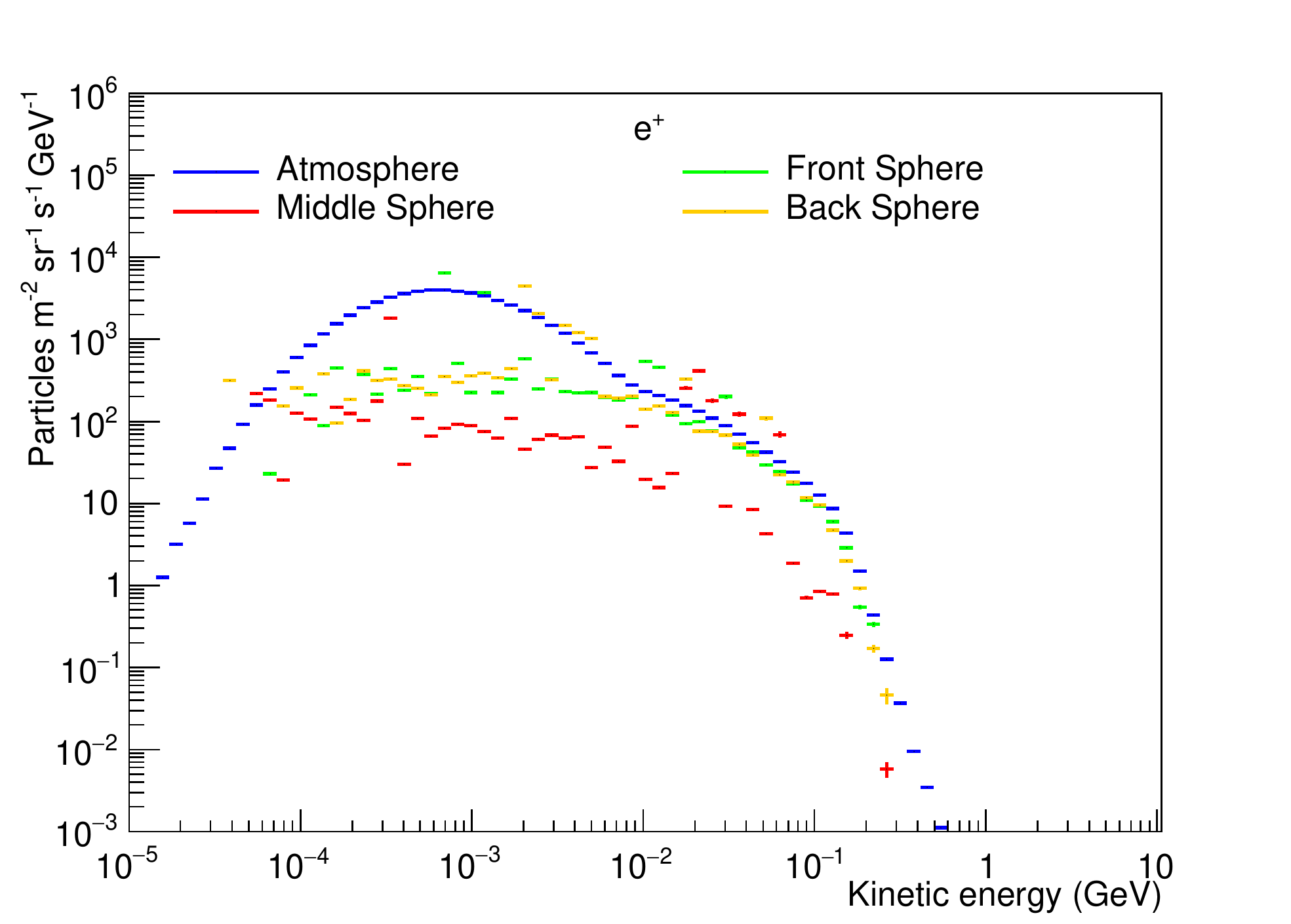}
\end{subfigure}
\begin{subfigure}[b]{0.49\textwidth}
\centering
\includegraphics[scale=0.3]{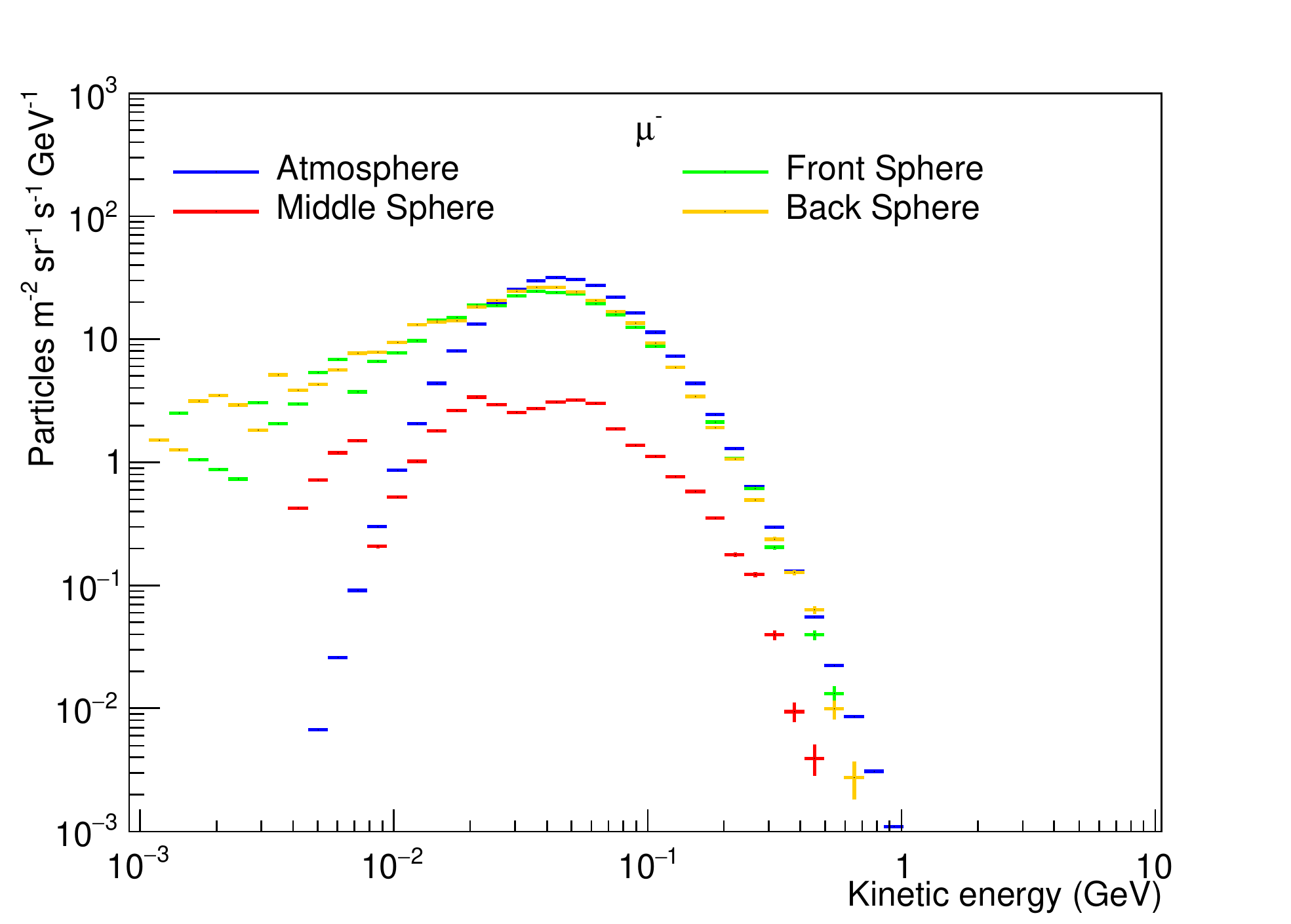}
\end{subfigure}
\begin{subfigure}[b]{0.49\textwidth}
\centering
\includegraphics[scale=0.3]{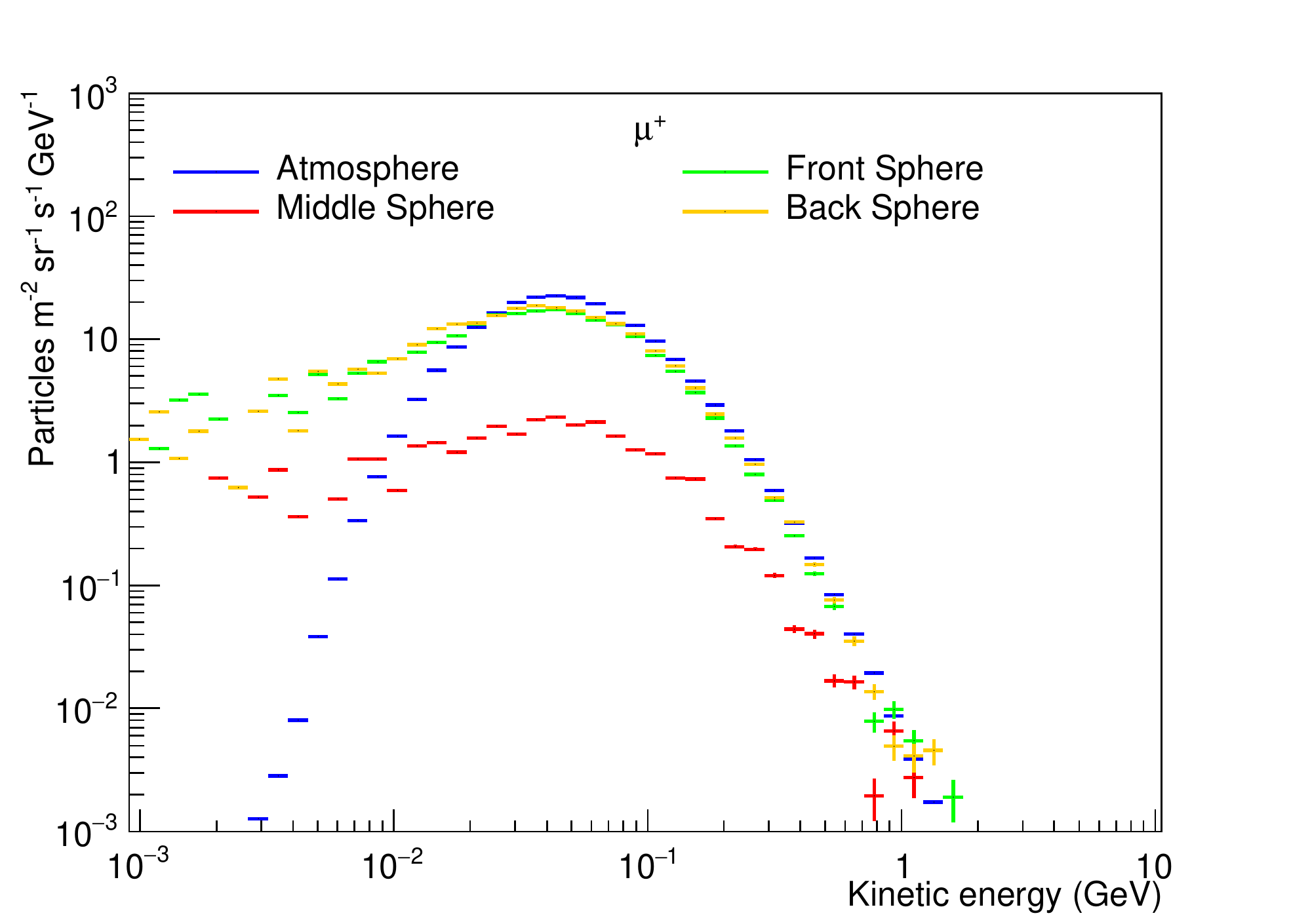}
\end{subfigure}
\caption{Particle flux for different upward going secondary GCR particles (proton, neutron,
$\gamma$, $e^{-}$, $e^{+}$, $\mu^{-}$, $\mu^{+}$) at aviation altitude and in four
different positions (outside:Atmosphere, front, middle and back side of the aircraft).}
\label{fig:radflup}
\end{figure}

The simulated flux of different secondary particles at three different positions
inside the aircraft, along with the flux outside the aircraft structure (i.e., in
the free atmosphere), for both downward and upward going particles can be seen
from Fig. \ref{fig:radfldn} and Fig. \ref{fig:radflup}. Here, we discarded the
downward going $\pi^{\pm}$ due to their negligible contribution to the total 
effective dose rate (see Table \ref{Tab:DoseT}). The change (in \%) of the integral
flux (particles $m^{-2} sr^{-1} s^{-1}$) for different individual secondary 
particles (as well as for the total contribution) in both downward and upward 
directions at different positions inside the fuselage with respect to the 
corresponding flux outside the aircraft, is given in Table \ref{tab:intfluxV}. 
The corresponding change (in $\mu$Sv/h) in effective dose rate in human body is 
tabulated in Table \ref{tab:doseV}. However, to calculate the effective dose 
rate at different positions inside the aircraft we considered the same 
fluence-to-dose-rate weighting factor obtained from the dose rate calculation 
in human phantom in the atmosphere as discussed in Sec. \ref{ssec:dose}. 

From these results, it is apparent that inside the aircraft fuselage the flux 
of downward going particles are increased in general (with exception for the 
muons). This may be attributed due to the secondary production by interaction 
of high energy particles with the aircraft materials have overwhelmed 
the absorption of the particles in the aircraft shell. On the other hand, a 
significant reduction of particle flux inside the fuselage for upward going 
particles is discernible from the results. Which is mainly due to the presence 
of the floor and central fuel tank. But, since downward particles contribute 
almost $\sim$96\% to the total effective dose rate, while upward particles 
contribute only $\sim$4\% (see Table \ref{Tab:DoseT}), the reduction of upward 
particle flux could not effectively reduce the dose rate. In fact this 
calculation shows about 9.7\% increase of total effective dose rate at the 
center of the fuselage compared to the free atmosphere, while effective dose 
rate due to neutrons increase by $\sim$ 6\%. The neutron dose rate measurement 
inside RB57-F military aircraft shows about 10\% increment in the radiation 
dose rate than outside the aircraft \cite{sing99}. In a previous calculation, 
\cite{ferr05} reported qualitatively similar findings of slight increment 
(depending on the position along the fuselage) of radiation dose in side the 
aircraft due to downward radiation, while the contribution from the upward 
radiation decremented due to the presence of the fuel and cargo content. 
However, the dose rate inside the fuselage may also change due to absorption 
of radiation by the internal structures, seats, passengers and other high-Z 
materials, which can be studied by MC simulation with detailed geometrical 
structure of the aircraft from the manufacture and corresponding measurements 
on board the aircraft.

\begin{table}
\centering
\caption{Percentage changes of integral flux (Particles $m^{-2} sr^{-1} s^{-1}$) for total
and individual different secondary GCR particles in front, middle and back section of the
fuselage with respect to the free atmosphere.}
%\begin{tabular}{ccccccc}
\begin{tabular}{C{0.15\textwidth}C{0.10\textwidth}C{0.10\textwidth}C{0.10\textwidth}C{0.10\textwidth}C{0.10\textwidth}C{0.10\textwidth}}
\hline
\multirow{2}{*}{Particles} & \multicolumn{2}{c}{Front (\%)} & \multicolumn{2}{c}{Middle (\%)} & \multicolumn{2}{c}{Back (\%)} \\ \cline{2-7}
          & Down & Up & Down & Up & Down & Up \\ \hline
$n$         & 04.51 & -11.53 & 	13.74 &    -86.87  &    05.59  &   -10.70  \\
$p$         & 00.28 & -49.61 & 	05.83 &    -88.52  & 	03.10  &   -40.39  \\
$\gamma$  & 04.14 & -30.87 & 	27.15 &    -88.20  &    04.59  &   -26.64  \\
$e^{-}$   & 12.25 & -78.56 & 	16.91 &    -94.03  &    13.85  &   -70.34  \\
$e^{+}$   & 16.37 & -42.29 & 	29.90 &    -67.18  &    18.06  &   -29.28  \\
$\mu^{-}$ & -2.74 & -15.86 & 	01.77 &    -88.64  &    -1.20  &   -11.59  \\
$\mu^{+}$ & -2.56 & -17.36 & 	01.02 &    -88.06  &    02.30  &   -11.02  \\
Total     & 04.60 & -29.58 & 	24.76 &    -88.09  &    05.26  &   -25.62  \\ \hline
\end{tabular}
\label{tab:intfluxV}
\end{table}

\begin{table}
\centering
\caption{Variation of effective dose rate for different downward and upward going
secondary GCR particles inside the aircraft fuselage.}
\begin{tabular}{C{0.15\textwidth}C{0.10\textwidth}C{0.10\textwidth}C{0.10\textwidth}C{0.10\textwidth}C{0.10\textwidth}C{0.10\textwidth}}
\hline
\multirow{2}{*}{Particles} & \multicolumn{2}{c}{Front ($\mu$Sv/h)} & \multicolumn{2}{c}{Middle ($\mu$Sv/h)} & \multicolumn{2}{c}{Back ($\mu$Sv/h)} \\ \cline{2-7}
          & Down & Up & Down & Up & Down & Up \\ \hline
$n$       &  0.0921	  & -0.0198	&	0.2808	&	-0.1493		& 0.1142 	& -0.0183 \\
$p$       &  0.0040		& -0.0042	&	0.0839	&	-0.0075		& 0.0446 	& -0.0034 \\
$\gamma$  &  0.0313		& -0.0075	&	0.2054	&	-0.0216		& 0.0347 	& -0.0065 \\
$e^{-}$   &  0.0453		& -0.0037	&	0.0626	&	-0.0045		& 0.0512 	& -0.0033 \\
$e^{+}$   &  0.0454		& -0.0015	&	0.0829	&	-0.0024		& 0.0501 	& -0.0010 \\
$\mu^{-}$ & -0.0044		& -0.0004	&	0.0028	&	-0.0027		& -0.001	& -0.0003 \\
$\mu^{+}$ & -0.0047		& -0.0004	&	0.0018	&	-0.0024		& 0.0042 	& -0.0003 \\ \hline 
\multirow{2}{*}{Total}    & 0.2091 &	-0.0379 &	0.7206 &	-0.1906 &	0.2973 &	-0.0334   \\ \cline{2-7}
& \multicolumn{2}{c}{0.1712} & \multicolumn{2}{c}{0.5299} & \multicolumn{2}{c}{0.2638} \\ \hline
\end{tabular}
\label{tab:doseV}
\end{table}

\begin{table}
\begin{small}
\centering
\caption{Dose rate as calculated from this simulation and using the dose conversion coefficient from \citet{pell00}.}
\label{Tab:doseComp}
\begin{tabular}{C{1.4cm}C{2.8cm}C{2.8cm}C{3.8cm}}
\hline
  Particles & Effective dose rate ($\mu Sv/h$)& Effective dose rate$^*$ ($\mu Sv/h$)& Ambient dose equivalent$^*$ ($\mu Sv/h$)\\
\hline
$n$       & 2.2159  & 2.0371 & 2.2398 \\
$p$       & 1.4486  & 2.1885 & 0.7790 \\
$\gamma$  & 0.7812  & 0.4644 & 0.1895 \\
$e^{-}$   & 0.3751  & 0.1924 & 0.1704 \\
$e^{+}$   & 0.2811  & 0.1454 & 0.2507 \\
$\mu^{-}$ & 0.1660  & 0.0952 & 0.0935 \\
$\mu^{+}$ & 0.1880  & 0.1154 & 0.1108 \\
$\pi^{-}$ & 0.0039  & 0.0023 & 0.0015 \\
$\pi^{+}$ & 0.0048  & 0.0072 & 0.0054 \\
\hline                
Total     & 5.4651  & 5.2484 & 3.8408 \\
\hline     
\end{tabular}
\footnotesize{$^{*}$ {using fluence-to-dose conversion coefficients for isotropic irradiation}}
\end{small}
\end{table}

\subsection{Comparison with previous dose calculations}
\label{ssec:comp}
This simulation yields that the total weighted sum of equivalent dose rate received by the 
female phantom is 5.72 $\mu$Sv/h, whereas that by the male phantom is 5.20 
$\mu$Sv/h. So the sex-averaged effective dose rate is 5.46 $\mu$Sv/h. The 
experimental or calculated dose rates received by the individual organs inside 
the human body (at aviation altitude), influenced by the aircraft structure at 
the geomagnetic latitude and altitude considered in this simulation are not 
available. So, we could not directly compare our simulated results for absorbed 
and equivalent dose rates on each organ of the male and female body. However,
we bench-marked our simulation at different stages of the calculation. 

\begin{figure}
\centering
\includegraphics[scale=0.38]{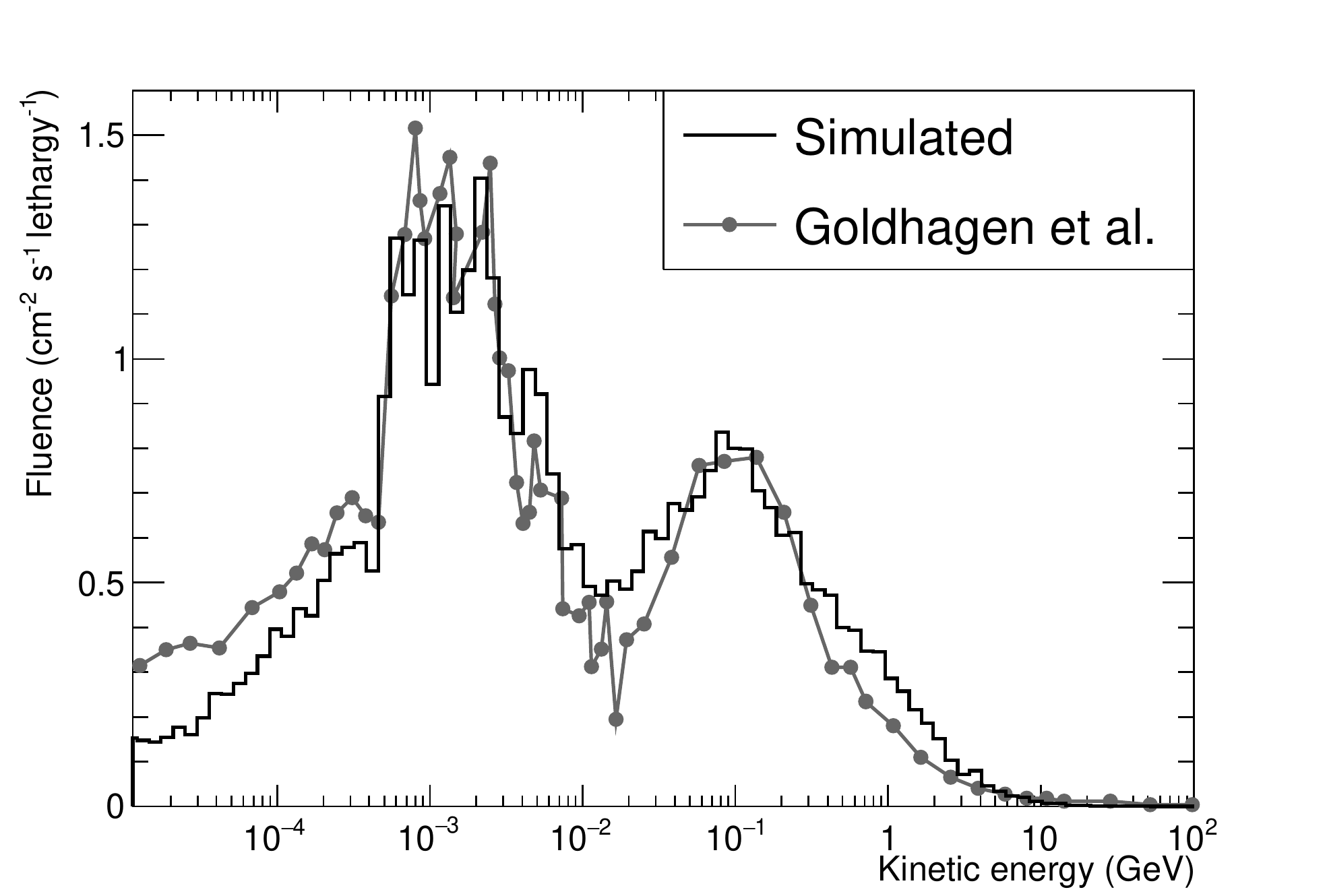}
\caption{Neutron fluence rate at 20 km altitude and in 1.0 - 1.1 rad 
geomagnetic latitude obtained in this simulation (black) and that measured 
by \citet{gold02} (gray).}
\label{fig:fluence}
\end{figure}

In Sec. \ref{ssec:fltalt}, we have already mentioned about the validation of 
the simulated proton flux at satellite altitude and atmospheric radiation flux 
at balloon altitude. We also compared the simulated neutron fluence rate at 
the aviation altitude with the experimental measurement reported by 
\citet{gold02}. Fig. \ref{fig:fluence} shows the neutron fluence rate per 
lethargy at 20 km altitude and in 1.0 - 1.1 rad geomagnetic latitude, as 
obtained in this simulation and compared to that measured by \citet{gold02} 
in the same latitude and altitude region. Considering this neutron fluence 
and conversion coefficients from \citet{pell00}, we calculated the effective 
dose rate (ambient dose equivalent rate) as 7.60 (9.12) $\mu$Sv/h which is 
within 10\% deviation of the measured value.

The radiation dose in the human phantom depends on many factors like: material 
distribution considered in the phantom, incident radiation spectra, energy 
range of the radiation, angular distribution of the incident flux etc. For 
example, \citet{sato09} showed the effect of using different source geometries
on radiation dose. Here, in this work, instead of using fluence to dose 
conversion coefficients, using micro-dosimetric technique we explicitly 
calculated the energy depositions in different organs for a wide range of 
radiation types and energy considering different angular distributions for 
downward and upward radiations. However, to compare our simulated dose rate, 
we also calculated the effective dose rate using the common conversion 
coefficients for isotropic irradiation from \citet{pell00} and the values 
for different radiation types are tabulated in Table \ref{Tab:doseComp}. 
The comparison shows that the total effective dose rate are within the 
acceptable uncertainty range, although some deviations in the effective dose 
rate for different particles are noticeable due to above mentioned factors.
 
Since, effective dose and equivalent dose are not measurable quantities, so 
ambient dose equivalent ($H^*(10)$) used instead for measurements. In Table 
\ref{Tab:doseComp} we also presented the $H^*(10)$ values for different 
particles calculated using corresponding conversion coefficients. The 
apparent underestimation of total ambient dose equivalent rate compared to 
the total effective dose rate can be understandable from the measurement of 
dose depth distribution aboard a series of flights from Cologne (GER) to 
Washington, DC (USA) using a quasi tissue-equivalent phantom sphere by 
\citet{vana03}. This measurement shows that the maximum dose equivalent 
appears in a depth of 50 to 60 mm rather than 10 mm, implying that $H^*(10)$ 
underestimates the actual whole-body radiation exposure. Similarly, 
\citet{peto10,vein11} showed that above few MeV, $H^*(10)$ can not provide 
a conservative estimate of the protection quantities. In fact, whether 
$H^*(10)$ saturates with energy, effective dose increases. \citet{ferr98} 
also conclude that the only up to a limited energy range $H^*(10)$ gives a 
conservative estimate of effective dose and for high energy neutrons.
% , their is no higher value of
% depth (d) in H$^*$(d), which can be used in operational quantity for
% radiation protection.
%%gives a conservative estimate of the effective dose in a limited neutron
%%energy range. Since the maximum of the dose equivalent distribution along
%%the principal diameter of a ICRU sphere gives a conservative approximation
%%of the limiting quantity only up to 100 MeV, no higher value of depth, d,
%%in H$^*$(d), can be specified for the definition of an operational quantity
%%appropriate for purposes of high energy neutron dosimetry in radiation
%%protection. 
For the validation of this simulation here we also compere
the calculated ambient dose equivalent rate (3.84 $\mu Sv/h$) in the simulated 
region ($\theta_{M}$ = 45$^\circ$--50$^\circ$, Altitude = 10 km, $\phi$ = 524 MV) with the
other reported values of ambient dose equivalent rate in \citet{ploc13,lind04,bott12,euro04}
which is determined by in-flight measurements through several radiation monitors and
in different altitude, latitude as well as solar condition. 
We simply interpolate the reported values for the comparison, which 
shows that our simulated dose rate is well within the acceptable limit of 30\%
uncertainty. 
%We also compared our simulated dose rate with other reported values in  
%\citet{bott12,euro04,lind04} (simply interpolating the reported values), which 
%shows that our simulated dose rate is well within the acceptable limit of 30\%
%uncertainty.

%%However, 
%%to validate our result, we compare the effective dose rate calculated in this 
%%work, with that calculated by Nowcast of Atmospheric Ionizing Radiation for 
%%Aviation Safety (NAIRAS) model \citep{nair13}. However, this model does not 
%%consider the influence of the aircraft structure to compute the dose rate. 
%%From this comparison, we saw that our simulated effective dose rate is well 
%%within the acceptable limit of 30\% to the effective dose rate as calculated 
%%by NAIRAS model (9.63 $\mu$Sv/h) which validate our result.

\section{Conclusions}
\label{sec:conc}
The gender specific radiation dose on the human body and its internal organs 
are calculated in this work through MC simulation, using most updated models 
describing the atmosphere and magnetosphere, CR flux, human phantoms and organ 
specific weighting factors. The calculation shows that, among all the incident 
primary particles, most of the radiation effective dose received by the human 
body is due to the neutrons, at the location of aviation altitude considered 
in this simulation. By interacting with the human phantoms secondary CRs also 
produce different cosmogenic radioactive nuclides with half lives ranging from 
less than a second to years. These radionuclides may give negligible 
contributions to the total effective dose received per hour during the flight, 
but can gradually increase with increasing altitude, latitude and exposure time 
and most importantly can have significant cumulative effect over the human 
lifespan. The calculation also reveals that the dose received by the human 
phantoms at flight altitude are mostly due to downward directed particles 
with a small contribution by the upward particles. Comparison of the calculated 
effective dose rate in this work with that calculated by various other computer 
codes and measurement as reported in \citet{ploc13,lind04,bott12,euro04} shows an agreement 
well within the acceptable error limit.
%from the NAIRAS model shows an 
%agreement well within the acceptable error limit.
Study of the shielding effect of the aircraft structure on the radiation dose 
is very important in aviation safety as this helps us to better predict 
radiation dose for radiological protection. By considering a detailed aircraft 
structure \citet{batt05} showed that, inside the aircraft, decrease of ambient 
dose equivalent is more significant than the decrease of effective dose compared 
to free atmosphere. Based on this they conclude that effective dose at the
free atmosphere can be used for individuals risk assessment. On the other hand
great care should be taken to consider the shielding effect on the ambient dose
equivalent when compared the calculated value with the measurement. Here, in 
this simulation, the consideration of the simplified aircraft model structure with 
only the outer shell, gives apparently odd result that the effective radiation 
dose is increased inside the aircraft. But this is understandable from the fact 
that the high-energy particle fluxes are converted to relatively low-energy 
particles by the shell material which subsequently contribute to the received 
dose by the human body more efficiently. However, this calculation can be 
improved by considering the the actual particle flux distribution produced 
inside the aircraft to calculate the energy depositions in the phantoms, rather 
than using the fluence-to-dose-rate weighting coefficients obtained from the 
calculation at the atmosphere outside the aircraft. Moreover, the calculation 
of radiation dose in human body inside the aircraft can further be improved 
by considering more realistic internal structure of the aircraft and passenger 
distribution, which is of course practically very challenging to implement in 
this kind of simulation.  

\section{Acknowledgment}
This work has been supported by the SERB (India) project EMR/2016/003870. We
also thank the Higher Education department of West Bengal, for a Grant-In-Aid
which allowed us to carry out the research activities at ICSP.

\biboptions{authoryear}
\bibliography{mybibfile}

\end{document}